\documentclass[journal,twocolumn,10pt]{IEEEtran}

\usepackage{cite}
\usepackage{amsmath,amssymb,amsfonts}
\usepackage{algorithmic}
\usepackage{algorithm}
\usepackage{graphicx}
\usepackage{textcomp}
\usepackage{xcolor}
\usepackage{subfigure}
\usepackage{graphicx}
\usepackage{booktabs}
\usepackage{multirow}
\usepackage{setspace}
\usepackage{stfloats}

\ifCLASSINFOpdf
\else
\fi

\begin{document}
 
\title{\huge Symbiotic Sensing and Communication: Framework and Beamforming Design}

\author{Fanghao Xia, Zesong Fei,~\IEEEmembership{Senior~Member,~IEEE}, Xinyi Wang,~\IEEEmembership{Member,~IEEE},  Weijie Yuan,~\IEEEmembership{Senior~Member,~IEEE}, \\ Qingqing Wu,~\IEEEmembership{Senior~Member,~IEEE}, Yuanwei Liu,~\IEEEmembership{Fellow,~IEEE}, and Tony Q. S. Quek,~\IEEEmembership{Fellow,~IEEE}
\vspace{-0.5cm}
	
	\thanks{Fanghao Xia, Zesong Fei, and Xinyi Wang are with the School of Information and Electronics, Beijing Institute of Technology, Beijing 100081, China (e-mail: xiafanghaoxfh@163.com, feizesong@bit.edu.cn, bit\_wangxy@163.com.).}
	
	\thanks{Weijie Yuan is with the Department of Electrical and Electronic Engineering, Southern University of Science and Technology, Shenzhen 518055, China (e-mail: yuanwj@sustech.edu.cn).}
	
	\thanks{Qingqing Wu is with the Department of Electronic Engineering, Shanghai Jiao Tong University, Shanghai 200240, China (e-mail: qingqingwu@sjtu.edu.cn).}
	
	\thanks{Yuanwei Liu is with the School of Electronic Engineering and Computer Science, Queen Mary University of London, London E1 4NS, U.K. (e-mail: yuanwei.liu@qmul.ac.uk).}
	
	\thanks{Tony Q. S. Quek is with the Information Systems Technology and Design Pillar, Singapore University of Technology and Design, Singapore 487372, Singapore (e-mail: tonyquek@sutd.edu.sg).}

}

\maketitle

\begin{abstract}
In this paper, we propose a novel symbiotic sensing and communication (SSAC) framework, comprising a base station (BS) and a passive sensing node. In particular, the BS transmits communication waveform to serve vehicle users (VUEs), while the sensing node is employed to execute sensing tasks based on the echoes in a bistatic manner, thereby avoiding the issue of self-interference. Besides the weak target of interest, the sensing node tracks VUEs and shares sensing results with BS to facilitate sensing-assisted beamforming. By considering both fully digital arrays and hybrid analog-digital (HAD) arrays, we investigate the beamforming design in the SSAC system. We first derive the Cramér-Rao lower bound (CRLB) of the two-dimensional angles of arrival estimation as the sensing metric. Next, we formulate an achievable sum rate maximization problem under the CRLB constraint, where the channel state information is reconstructed based on the sensing results. Then, we propose two penalty dual decomposition (PDD)-based alternating algorithms for fully digital and HAD arrays, respectively. Simulation results demonstrate that the proposed algorithms can achieve an outstanding data rate with effective localization capability for both VUEs and the weak target. In particular, the HAD beamforming design exhibits remarkable performance gain compared to conventional schemes, especially with fewer radio frequency chains.

\end{abstract}


\IEEEpeerreviewmaketitle

\section{Introduction}

The sixth generation (6G) mobile communication network is expected to be a wide-coverage and intelligent system, providing both high-quality wireless connectivity and high-accuracy sensing \cite{8999605,9625159}. With separate development for decades, both radar and communication systems exhibit outstanding performance and have been applied in various fields. However, the separate operation of these two functionalities within physically isolated sub-systems results in low spectral efficiency and inter-system interference, especially with the explosive growth in the number of devices. Fortunately, multi-input-multi-output (MIMO) radar and massive MIMO communication were successively proposed and developed towards millimeter wave (mmWave) frequency bands and larger antenna arrays \cite{5419124,6824752}, resulting in increasing similarities among sensing and communication systems in terms of hardware architectures, channel characteristics, and signal processing \cite{9737357}, and consequently offering promising opportunity for integrated sensing and communications (ISAC).


Generally, according to the integration level, ISAC can be divided into two types, i.e., (a) radar and communication co-existence (RCC) and (b) dual-functional radar-communication (DFRC). The RCC system aims at eliminating the interference between the radar and communication system with priorly shared knowledge. It thus can be implemented with almost no change in either the hardware or the system architecture. One straightforward realization is to allocate orthogonal time or frequency resources to sensing and communication systems. For example, the communication system in \cite{6331681} was allowed to transmit the signals with the time and frequency resource not occupied by the radar to prevent collisions. Nevertheless, the communication performance is restricted by the radar system, and the two systems cannot operate simultaneously. Comparatively, spatial-division waveform design has drawn significant research. By applying mmWave and MIMO techniques, the communication and sensing waveform can be transmitted over different orthogonal spatial beams in sparse channels \cite{8871348}. In \cite{6503914}, the radar waveform was projected onto the null space of the interference channel to eliminate interference for the communication system, while the radar interference can be reduced by using spatial filter \cite{6558027}. 

In the aforementioned works, the system design is unilaterally from either the radar or communication perspective, which focuses on introducing another system while not sacrificing the performance of the original system. As a step further, the communication and radar systems can be jointly designed by simultaneously considering performance metrics for both systems. In \cite{7470514}, the radar beamforming matrix and communication covariance matrix were jointly optimized to minimize the effective interference power of radar while maintaining average capacity. The robust beamforming design was considered in \cite{7898445} by maximizing the radar detection probability while imposing an appropriate constraint on the SINR at users' receivers. \textcolor{black}{For the unbiased (or asymptotically unbiased) parameter in radar sensing, the Cramér-Rao lower bound (CRLB) provides a lower bound mean square error. The authors in \cite{8892631} minimized CRLB for the angles of arrival (AoA) estimation with MIMO radar, while guaranteeing communication quality of the multiple-input single-output (MISO) communication system, which validated the effectiveness of utilizing the CRLB as the sensing performance metric.} In addition, both radar-centric and communication-centric designs were considered in \cite{8834831}.

The DFRC design, with higher integration and lower hardware costs, has recently attracted widespread interest. Specifically, DFRC systems transmit integrated signals to simultaneously perform sensing and communication functionalities. In \cite{8288677}, the authors studied the DFRC system in a MISO scenario, where the beamformer was designed such that the obtained beampattern approaches the desired radar beampattern while satisfying the users' SINR constraint. In \cite{8386661}, the power of multi-user interference (MUI) was minimized, where the beampattern mismatch level was considered to achieve a trade-off between communication and sensing performance. The author in \cite{9416177} employed reconfigurable intelligent surface (RIS) to assist the DFRC system. The introduced extra degrees of freedom (DoFs) facilitate the system to improve the sum rate under the sensing beampattern constraint. Furthermore, the CRLB was also taken as a sensing performance metric in the DFRC system. In \cite{9652071}, the CRLB optimization framework for the point target and the extended target was proposed, while guaranteeing QoS of communication. By contrast, \cite{10364735} aimed at maximizing the achievable sum rate of the multi-user under the CRLB constraint for parameter estimation. The fundamental CRLB-rate trade-off was also investigated in multi-antenna multi-target scenario \cite{10251151}. 

\textcolor{black}{The aforementioned works all focused on the architecture of full radio frequency (RF) chains, which include signal mixers and analog-to-digital converters, and are comparable in number to the antenna elements. However, the prohibitive cost makes digital beamforming infeasible for large-scale antenna arrays in mmWave systems.} To reduce the hardware cost of substantial RF chains, the DFRC systems equipped with hybrid analog-digital (HAD) array were investigated, where the massive antennas are connected with a small number of RF chains through analog phase shifters \cite{1519678}. The researches include  near-field effect \cite{10135096}, sum-rate maximization \cite{9366836}, MUI minimization \cite{liyanaarachchi2021joint}, SINR maximization \cite{9538857}, CRLB minimization \cite{9868348}, and performance trade-off \cite{9557817}.

Despite the above research progress, the aforementioned RCC system and DFRC system exhibit inherent limitations. In the case of the RCC system, the communication and radar systems remain physically segregated, necessitating dual sets of independent transceivers and consequently resulting in elevated hardware expenditures \cite{8288677}. Moreover, the communication and radar systems actively transmit their own signals, leading to energy wastage and mutual interference \cite{10024901}. The interference cancellation requires a sophisticated beamforming design and suffers from performance degradation due to the limited DoFs \cite{9416177}. Although the above issues do not exist in DFRC systems, self-interference is a significant hurdle that cannot be bypassed. As stated in \cite{9296833}, without modifying the current half-duplex hardware architecture, the leakage of transmitted signals can easily overwhelm the reflected echoes. Although plenty of full-duplex techniques have been studied in \cite{6832464,9724187}, they cannot support simultaneous high-throughput communication and high-accuracy remote sensing, let alone the extremely high cost of replacing the hardware of massive existing BSs. 

\textcolor{black}{To address this issue, the bistatic ISAC system model was proposed to avoid the strong self-interference in a monostatic ISAC system by employing a pair of physically separated sensing transceivers and maintaining the merit of co-designing radar sensing and communications on shared spectrum and hardware \cite{jiao2023information}. In \cite{falcone2010experimental}, the authors have demonstrated the feasibility of bistatic sensing based on passive radar processing of communication signals. In \cite{9860521}, the estimation of the AoA, delay, and Doppler shift was achieved for a cooperative bistatic ISAC system. A deep-learning-based AoA estimation algorithm was proposed to improve the sensing performance in \cite{10464930}. In \cite{brunner2024bistatic}, the sensing node is connected with BS via a wireless link, where the ISAC signals undergo synchronization and communication processing to reconstruct the original waveform at the sensing node, which is then utilized for target sensing.}

Motivated by the above, in this paper, we propose to deploy a sensing node as a mere receiver to enable the passive localization of targets based on echoes of the communication signals transmitted by BS. While the communication signals transmitted by the BS are exploited as the probing signal for target parameter estimation, it can also help to localize the vehicle users (VUEs). \textcolor{black}{Different from the bistatic ISAC system \cite{jiao2023information,falcone2010experimental,9860521,10464930,brunner2024bistatic}, where only the communication-assisted sensing was considered, the sensing results are also leveraged to facilitate communication beamforming design in this paper. Specifically, based on the localization results from the sensing node, the BS can reconstruct LoS components of VUEs and perform adaptive beamforming, hence saving the pilot overhead for beam tracking. Through such coordination, the sensing node and communication BS constitute a symbiotic sensing and communication (SSAC) system.} In the considered SSAC system, it is of vital importance to achieve high-accuracy estimation of two-dimensional (2D) AoAs for both VUEs and the target. Although there have been numerous works \cite{9366836,9947033,9685274,10304580} studying beamforming in ISAC systems, they cannot be straightforwardly used in SSAC systems due to the bistatic sensing manner. Therefore, we investigate the derivation of CRLB for the estimation of 2D AoAs and beamforming design for fully digital and HAD arrays.

Our main contributions are summarized as follows.
\begin{figure}[t]
	\centering
	\includegraphics[width=1\linewidth]{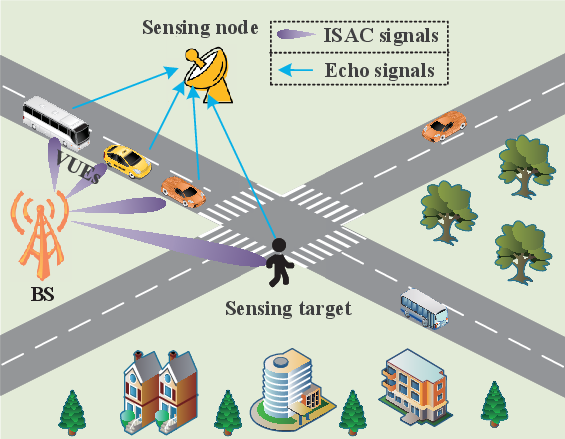}
	\caption{An illustration of the considered SSAC system.}
	\label{systemmodel}
	\vspace{-0.5cm}
\end{figure}

\begin{itemize}
\item We formulate a novel SSAC framework as shown in \figurename{ \ref{systemmodel}}, where a passive sensing node is deployed to sense VUEs and targets based on the echoes of communication signals transmitted by BS, while the BS leverages the sensing results shared by sensing node to assist beamforming design. Based on this setup, we derive the CRLB for estimating the 2D AoAs of the target. Then, we formulate an achievable sum rate maximization problem by designing the beamformer for both fully digital arrays and HAD arrays subject to the constraints of the CRLB to ensure the AoA estimation accuracy.

\item For the fully digital architecture, we propose a penalty dual decomposition (PDD)-based framework to solve the formulated problem. We first transform the objective function into a more tractable form via the fractional programming (FP) technique. Then, the PDD framework is employed to address the non-convex CRLB constraint, and decompose the problem into several sub-problems, with the gradient projection algorithm used to solve the beamforming optimization sub-problem.

\item We extend the proposed algorithm to the HAD beamforming design considering both fully-connected architecture and partially-connected architecture, where the digital beamformer and the analog beamformer are optimized iteratively. The digital beamformer is optimized by the gradient projection method, while for the analog beamformer design, we propose a manifold optimization-based algorithm to address the unit-modulus constraints. In particular, the extra penalty method is leveraged to address the power budget constraint for the fully-connected architecture, and the objective function is transformed to reduce the computational complexity for the partially-connected architecture.

\item Simulation results validate the effectiveness of the proposed algorithms for fully digital arrays and HAD arrays. The results show that the communication performance obtained by the proposed algorithm outperforms that of the conventional beamforming design scheme, while the sensing performance is guaranteed. In addition, the trade-off between communication and sensing performance is also revealed.
\end{itemize}

The remainder of this paper is organized as follows. Section II introduces the SSAC system model and the signal model. Section III derives the CRLB of AoAs and presents the problem formulation. The PDD-based fully digital beamformer optimization algorithm is proposed in Section IV. The alternating optimization of the HAD array is described in Section V. Then, simulation results are presented in Section V. Finally, Section VI concludes this paper.

\textit{Notations}: $a$, $\bf{a}$, $\bf{A}$ denote complex scalar value, vector, and matrix, respectively; ${\left[  \cdot  \right]^*}$, ${\left[  \cdot  \right]^T}$, and ${\left[  \cdot  \right]^H}$ denote the conjugate, transpose, and conjugate-transpose operations, respectively; $\left| {\cdot} \right|$ and $\left\|  \cdot  \right\|_F$ denotes the absolute value and the Frobenius norm, respectively; $\mathbb{C}$ represent the set of complex number; ${\rm{diag}}\left(  \cdot  \right)$ refers to the diagonal operation; \textcolor{black}{$\Re \left\{ \cdot \right\}$ and $\Im \left\{ \cdot \right\}$ denote the real part and imaginary part of its argument, respectively}; ${\bf{I}}_n$ denotes the identity matrix with the dimension of $n\times n$;  $\mathcal{C} \mathcal{N}\left(\mu, \sigma^{2} \right)$ denote the complex Gaussian distributions with mean $\mu$ and variance $\sigma^{2}$; ${\bf{A}} \otimes {\bf{B}}$ represents the Kronecker product of ${\bf{A}}$ and ${\bf{B}}$; ${\bf{A}} \odot {\bf{B}}$ represents the Hadamard product of ${\bf{A}}$ and ${\bf{B}}$; \textcolor{black}{$[\cdot]^+$ represents $\max \{\cdot, 0\}$}.

\section{System Model}	
As shown in \figurename{ \ref{systemmodel}}, we consider an SSAC system, with the vehicle-to-everything (V2X) scenario taken as an example, where an $M$-antenna BS serves $K$ single-antenna VUEs in the downlink while collaborating with a passive sensing node to achieve sensing for a target of interest, \textcolor{black}{which can be a suspicious human, a pet animal, etc. In general, the radar cross-section (RCS) of other moving targets is much less than that of vehicles in the V2X scenario, making it difficult for radar sensing. For clarity, the target of interest is hereinafter referred to as the weak target.} The set of the VUEs is denoted as $\mathcal{K} = \{1,2, \cdots, K\}$. The sensing node is equipped with a uniform planar array (UPA) with $N = N_h \times N_v$ antennas for receiving echoes, where $N_h$ and $N_v$ represent the number of antennas in horizontal and vertical directions, respectively. \textcolor{black}{The BS and the sensing node are connected with an optical cable for clock synchronization and synchronous data exchange in real time \cite{4726071,5360722}}. \textcolor{black}{The locations of the BS and the sensing node are assumed to be fixed and known in advance.} By receiving the echoes of communication signals, not only the vehicle can be located to construct channel state information (CSI) for communication enhancement, but also the target can be detected to avoid a collision. 
\subsection{Communication Signal Model}
By assigning a dedicated beamforming vector for each VUE, the complex baseband transmitted signal at the BS can be expressed as $ {\bf x}(t) = \sum\nolimits_{k = 1}^K {{\bf w}_k s_k(t) }$, where $s_k(t)$  denotes the transmitted data for user $k$ at time slot $t$, which is assumed as independent Gaussian random signal with zero mean and unit power, and its covariance matrix is expressed as $ {\bf R}_s = \mathbb{E}\{{\bf s}(t){\bf s}(t)^H\} = {\bf I}_k$, ${\bf{s}}(t) = {\left[ {{s_1}(t), \cdots,{s_K}(t)} \right]^T}$.  ${\bf W} = \left[{\bf w}_1,{\bf w}_2, \cdots, {\bf w}_K \right] \in \mathbb{C}^{M \times K}$ denotes beamforming matrix. Note that for the HAD array, the beamforming matrix can be expressed as ${\bf W} = {\bf F}_a{\bf F}_{d}$, where ${\bf F}_a$ and ${\bf F}_{d}$ denote the analog beamforming matrix and the digital beamforming matrix, respectively.  Thus, the signal received by the $k$-th VUE is expressed as 
\begin{align}
	\label{eq2.1}
	{y_k}(t) = {\bf h}_k^H{ {\bf w}_k}{s_{ k}}(t) + \sum\limits_{j \ne k} {{\bf h}_k^H{{\bf w}_j}{s_{j}}(t)}  + { n}_c,
\end{align}
where ${{n}}_c \sim \mathcal{C} \mathcal{N}\left(0, \sigma_c^{2} \right)$ denotes the additive white Gaussian noise at the $k$-th VUE, ${\bf h}_k \in \mathbb{C}^{M \times 1}$ denotes the baseband equivalent channels from BS to VUE $k$. \textcolor{black}{In this paper, we consider mmWave channels based on Saleh–Valenzuela (S-V) channel model \cite{1146527}, which consists of both line-of-sight (LoS) and non-LoS paths, 
\begin{align}
	\label{eq2.2}
	{\bf h}_k = \sqrt{\rho_0 d_{bk}^{-\alpha_0}} \left({\bf{b}}\left( {{\psi _k}} \right) + \sum\limits_{i = 1}^{{n_{NL}}} {{\alpha _{n,i}}{\bf{b}}\left( {{\psi _{n,i}}} \right)} \right),
\end{align}
where ${\rho _0}$ denotes the channel gain at the reference distance of 1 meter,  $d_{bk}$ denotes the distance between the BS and VUE $k$, $\alpha_0$ denotes the path loss exponent, $\psi_k$ denote the angle of departure (AoD), i.e., the angle of VUE $k$ respect to the BS, $n_{NL}$ denotes the number of NLoS paths, $\alpha _{n,i}$ and $ \psi _{n,i} $ denotes the small-scale fading and angle of the corresponding path,} \textcolor{black}{${\bf{b}} ( \cdot )$ denotes the transmit steering vector of the BS, i.e., 
\begin{align}
	\label{eq2.3}
	{\bf{b}}\left( \psi  \right) = {\left[ {1,{e^{j\frac{{2\pi {d_c}}}{\lambda_c }\sin (\psi )}}, \cdots ,{e^{j\frac{{2\pi {d_c}}}{\lambda_c }(M - 1)\sin (\psi )}}} \right]^T},
\end{align} 
where $\lambda_c$ denotes the carrier wavelength, and $d_c$ denotes the antenna spacing that is set as $d_c = 0.5 \lambda_c$ in general.}

Based on \eqref{eq2.1}, the SINR for VUE $k$ and the achievable sum rate are given as 
\begin{align}
	\label{eq2.4}
	{\gamma _k} = \frac{{{{\left| {{\bf h}_k^H{{\bf w}_k}} \right|}^2}}}{{\sum\limits_{j \ne k} {{{\left| {{\bf h}_k^H{{\bf w}_j}} \right|}^2}}  + \sigma _c^2}},
\end{align} 
\begin{align}
	\label{eq2.5}
	R_{\rm sum} = \sum\limits_{k = 1}^K {\log_2 (1 + {\gamma _{k}})}.
\end{align}
\subsection{Sensing Signal Model}
\textcolor{black}{In this paper, both VUEs and the target are modeled as an unstructured point.} \textcolor{black}{The BS and the sensing node perform radar sensing following a bistatic mode based on the echo signals from VUEs and target. Before the AoA estimation, the fast Fourier transform (FFT) and constant false alarm rate (CFAR) are executed to extract the Doppler information of the vehicle users and target. The impact of Doppler frequency can then be compensated effectively at receivers. After the compensation, the received echoes at the sensing node can be modeled as}
\begin{align}
	\label{eq2.6}
	{\bf Y}_r = \sum\limits_{k =1}^K {{{\bf H}_{s,k}}{\bf X}}  + {{\bf H}_{s,t}} {\bf X} + \sum\limits_{c = 1}^C {{\alpha _c}{\bf{b}}\left( {{\psi _c}} \right){{\bf{b}}^H}\left( {{\psi _c}} \right)}  + {\bf N}_s,
\end{align} 
where ${\bf N}_s \in \mathbb{C}^{N\times M}$ denotes the noise at the sensing node with each entry following $\mathcal{C} \mathcal{N}\left(0, \sigma_r^{2} \right)$, ${\bf X} \in \mathbb{C}^{M\times L} = [{\bf x}(1), \cdots, {\bf x}(L)]$ denotes the transmitted signals in a Range-Doppler bin with the length being  $L$, \textcolor{black}{$C$ denotes the number of scatters, $\alpha_c$ denotes the reflection coefficients of the corresponding scatter, the first three terms are the echoes from $K$ VUEs, the target and scatters\footnote{\textcolor{black}{We assume that the interference signals from the BS and clutters can be eliminated based on the Doppler frequency difference via moving target indicators-moving target detection (MTI-MTD) technique \cite{10392309} and AoA difference via Space-Time Adaptive Processing (STAP) technique \cite{845251}. Note that the signal reflected by the moving scatters cannot be eliminated; thus, they should also be viewed as sensing targets to be tracked for driving safety.}}, respectively, } ${\bf H}_{s,k}$ and ${{\bf H}_{s,t}}$ denote the BS-VUE $k$-sensing node link and BS-target-sensing node link, which are modeled as
\begin{subequations}
	\begin{align}
	{\bf H}_{s,k} &=  e^{j\varphi_k} \sqrt{\rho_0 \zeta_k d_{bk}^{-2} d_{kr}^{-2}}{\bf{a}}_r\left( { \theta_k}, { \phi_k} \right)   {\bf{b}}^H \left( {{ \psi_k}} \right),   \label{eq2.7a}\\
	{\bf H}_{s,t} &=  e^{j\varphi_t} \sqrt{\rho_0 \zeta_t d_{bt}^{-2} d_{tr}^{-2}} {\bf{a}}_r\left( { \theta_t}, { \phi_t} \right)   {\bf{b}}^H \left( {{ \psi_t}} \right),\label{eq2.7b} 
	\end{align}
\end{subequations}
where $e^{j\varphi_k}$ and $e^{j\varphi_t}$ denote the reflection coefficient phase shift of $k$-th VUE and the target, respectively,  $\zeta_k$ and $\zeta_t$ denote the corresponding reflection coefficient amplitude, i.e., RCS, $ d_{bt} $, $ d_{kr} $, and $ d_{tr} $ denote the distance between the BS and target, between the $k$-th VUE to the sensing node, and between the target and sensing node, respectively, $\psi_t$ denote the angle of departure (AoD), i.e., the angle of target with respect to the BS, $\theta_k$ ($\theta_t$) and $\phi_k$ ($\phi_t$) denote the horizontal and vertical angles of arrival (AoAs), respectively. \textcolor{black}{ $ {\bf{a}}_r\left( { \theta}, { \phi} \right) = {\bf{a}}_{rh}\left( { \theta}, { \phi} \right) \otimes {\bf{a}}_{rv}\left( { \phi} \right) $
is the array response vector at the sensing node, where
\begin{subequations}
	\begin{align}
		&{{\bf a}_{rh}}(\theta ,\phi ) \label{eq2.8a}\\
		&=  \left[ {1,{e^{\frac{{2\pi {d_c}}}{\lambda_c } \sin (\theta )\sin (\phi )}}, \cdots ,{e^{\frac{{2\pi {d_c}}}{\lambda_c } ({N_h} - 1)\sin (\theta )\sin (\phi )}}} \right]^T, \notag \\
		&{{\bf a}_{rv}}(\phi ) = \left[ {1,{e^{\frac{{2\pi {d_c}}}{\lambda_c } \cos (\phi )}}, \cdots ,{e^{\frac{{2\pi {d_c}}}{\lambda_c } ({N_v} - 1)\cos (\phi )}}} \right]^T. \label{eq2.8b} 
	\end{align}
\end{subequations}
}

For sensing, we are interested in estimating the 2D AoAs $\theta$ and $\phi$ for both VUEs and the target. \textcolor{black}{Leveraging the prior information that the vehicles and the target are located on the ground, the VUEs and the target can be localized based on the 2D AoAs by taking the sensing node as an anchor. Specifically, the three-dimensional Cartesian coordinates of the sensing node and the target (or a vehicle) are expressed as $L_s = [x_s,y_s,h_s]^T$ and $L_t = [x_t,y_t,h_t]^T$, respectively. Denote $\Delta D = L_s-L_t = [d_x,d_y,d_h]^T$, which can be solved as $\|\Delta D\| = \frac{{{d_h}}}{{\cos \phi }}$, ${d_x} = \|\Delta D\|\sin \theta$, ${d_y} = \sqrt {{{\left( \|{\Delta D}\| \right)}^2} - d_x^2 - d_h^2}$, where $d_h$ can be obtained in prior. Then, the target (or a vehicle) can be located.}

In general, the RCS of VUEs is relatively large, and the echo power of VUEs is strong enough to achieve 2D AoAs estimation via the two-dimensional multiple signal classification technique (2D-MUSIC) \cite{1143830}. \textcolor{black}{The sensing results are fed back to the BS via the optical cable. Based on the position information, the dominated LoS components of the VUEs' mmWave channels can be reconstructed at the BS via \eqref{eq2.2} \cite{9947033,9685274,10304580}, which is leveraged to design the fully digital beamformer or HAD beamformer.} \textcolor{black}{In this paper, we focus on the scenario that sensing information is updated in real-time based on the echo signals. Additionally, the issue of sensing result obsolescence can be addressed using existing predicting methods, such as the extended Kalman filter \cite{9171304}, factor graph \cite{9246715}, and deep learning \cite{10304580}.}

The characteristics of the target are unpredictable. Especially when the RCS is small, it is challenging to accurately position the target, which poses a serious threat to road safety. To achieve accurate sensing, the sensing node first processes echo signals using spatial filtering to create nulls in the beampattern corresponding to the AoAs of the VUEs \cite{van2002optimum}, and then estimates the target's parameters via the proposed reduced-dimension maximum likelihood estimation (RD-MLE), which is presented in Appendix A.


\section{Problem Formulation}
In this paper, we aim to maximize the achievable sum rate of VUEs while guaranteeing the sensing performance for the weak target. In the following, we first derive the CRLB of 2D AoAs and then formulate the corresponding optimization problem.

\subsection{CRLB Derivation}

\begin{figure*}[b]	
	\hrulefill
	\begin{align}
		\label{eq2.11}
		&\frac{{\partial {\bf y}_0 }}{{\partial \boldsymbol{\xi} }} = \left[ { {\rm vec}\left( {{\alpha}\dot {\bf a}_r^{{\theta _t}} {{\bf b}^H} {\bf X}} \right),    {\rm vec}\left( {{\alpha}\dot {\bf a}_r^{{\phi_t}}{{\bf b}^H} {\bf X}} \right), {\rm vec}\left( {{\alpha} {\bf a}_r{ \dot {\bf b}}^H {\bf X}} \right) , {\rm vec}\left( {{\bf a}_r{{\bf b}}^H {\bf X}} \right),{\rm vec}\left( j{{\bf a}_r{{\bf b}}^H {\bf X}} \right) } \right], \tag{12} \\
		\label{eq2.12}
		&\dot {\bf a}_r^{{\phi _t}} = \frac{\partial {\bf a}_r \left( \theta_t, \phi_t \right)} { \partial \phi_t}  = \left( {j\pi \sin \theta_t \cos \phi_t {{\left[ {0,1, \cdots ,{N_h} - 1} \right]}^T} \odot {{\bf a}_{rh}}(\theta_t ,\phi_t )} \right) \otimes { {\bf a}_{rv}}(\phi_t ) \notag \\
		& \qquad \qquad \qquad \qquad \quad + {a_{rh}}(\theta ,\phi ) \otimes \left( { - j\pi \sin \phi {{\left[ {0,1, \cdots ,{M_v} - 1} \right]}^T} \odot {a_{rv}}(\phi )} \right),\tag{13}\\
		\label{eq2.13}
		&\dot {\bf a}_r^{{\theta _t}} = \frac{\partial {\bf a}_r \left( \theta_t, \phi_t \right)} { \partial \theta_t}  = \left( {j\pi \cos \theta_t \sin \phi_t {{\left[ {0,1, \cdots ,{N_h} - 1} \right]}^T} \odot {{\bf a}_{rh}}(\theta_t ,\phi_t )} \right) \otimes { {\bf a}_{rv}}(\phi_t ), \tag{14}\\
		\label{eq2.14}
		&\dot {\bf b} = \frac{\partial {\bf b} \left( \psi_t \right)} { \partial \psi_t}  =  {j\pi \cos \psi_t {{\left[ {0,1, \cdots ,{M} - 1} \right]}^T} \odot {\bf b} \left( \psi_t \right)}, \tag{15}
	\end{align} 
\end{figure*}

To evaluate the AoA estimation performance, we derive the CRLB as the sensing metric. First, we vectorize the received signals after spatial filtering as follows
\begin{align}
	\label{eq2.9}
	{\bf y}_r = {\bf y}_0 + {\bf n}_r,
\end{align} 
where ${\bf y}_0 = \alpha {\bf g}(\theta_t,\phi_t,\psi_t)$, $\alpha = e^{j\varphi_t} \sqrt{\rho_0 \zeta_t d_{bt}^{-2} d_{kt}^{-2}} $ denotes the complex reflection coefficient, ${\bf g}(\theta_t,\phi_t, \psi_t)  = {\rm{vec} } \left( {\bf{a}}_r\left( { \theta_t}, { \phi_t} \right)   {\bf{	b}}^H \left( {{ \psi_t}} \right) \bf X \right) $ with the dimension of $NL \times 1$, ${\bf n}_r = {\rm vec} \left({\bf N}_r\right)$. It can be observed that the ${\bf y}_r$ is an Gaussian observation, i.e., ${\bf y}_r \sim \mathcal{C} \mathcal{N}\left(\alpha {\bf{g}}, {\bf R}_n \right)$, where  $ {\bf R}_n = \sigma_r^{2} {\bf I}_{NL}$. The conditional probability density function (PDF) of ${\bf y}_r$ with given $\boldsymbol{\xi}$ is expressed as
\begin{align}
	\label{eq2.9add}
	p\left( {{{\bf{y}}_r}|{\boldsymbol{\xi}} } \right) = \frac{1}{{{\pi ^{NL}}\det \left( {{{\bf{R}}_n}} \right)}}\exp \left( { - {{\left( {{{\bf{y}}_r} - {{\bf{y}}_0}} \right)}^H}{\bf{R}}_n^{ - 1}\left( {{{\bf{y}}_r} - {{\bf{y}}_0}} \right)} \right),
\end{align} 
where ${\boldsymbol{\xi}} = \left[\theta_t, \phi_t, \psi_t, \Re(\alpha), \Im(\alpha) \right]^T$ denotes the unknown parameters to be estimated. \textcolor{black}{$\Re(\alpha)$ and $\Im(\alpha)$ denotes the real part and imaginary part of the complex reflection coefficient $\alpha$.} According to \cite{kay1993fundamentals}, the Fisher information matrix (FIM) for estimating the vector ${\boldsymbol{\xi}}$ is given as \textcolor{black}{
\begin{align}
	\label{eq2.10}
		{\bf J_{\xi}}	&= 2 \Re \left\{ {  \frac{{\partial {{\bf y}_0^H}}}{{{{ \partial \boldsymbol{\xi}} }}}  {\bf R}_n^{ - 1} \frac{{\partial {{\bf y}_0}}}{{{{\partial \boldsymbol{\xi}} }}}   } \right\}  + {\rm Tr}\left\{ {{\bf R}_n^{ - 1}\frac{{\partial {{\bf R}_n}}}{\partial \boldsymbol{\xi}} {\bf R}_n^{ - 1}\frac{{\partial {{\bf R}_n}}}{\partial \boldsymbol{\xi}}} \right\} \notag \\
& = \frac{2}{{\sigma _n^2}} 2 \Re \left\{ {  \frac{{\partial {{\bf y}_0^H}}}{{{{\partial \boldsymbol{\xi}} }}}  \frac{{\partial {{\bf y}_0}}}{{{{\partial \boldsymbol{\xi}} }}}  } \right\},
\end{align}
where $\frac{{\partial {{\bf y}_0}}}{\partial \boldsymbol{\xi}} $ is given as \eqref{eq2.11}-\eqref{eq2.14} at the bottom of this page.} \addtocounter{equation}{4} By denoting the blocked matrix 
\begin{equation}	
	\label{eq2.15}	
	\begin{aligned}	
		{\bf \Xi}	&= \left[ {\bf \Xi}_{\theta_t}, {\bf \Xi}_{\phi_t}, {\bf \Xi}_{\psi_t}, {\bf \Xi}_{\Re (\alpha)},{\bf \Xi}_{\Im(\alpha)} \right] \\
		&= [{\alpha}\dot {\bf a}_r^{{\theta _t}}{ {\bf b}^H}, {\alpha}\dot {\bf a}_r^{{\phi _t}}{ {\bf b}^H},{{\alpha} {\bf a}_r{ \dot {\bf b}}^H} ,{\bf a}_r{ {\bf b}^H}, j{\bf a}_r{ {\bf b}^H}],
	\end{aligned} 
\end{equation}
\textcolor{black}{the element at the $\ell$-th row and the $q$-th column of ${ \left[{{\bf J_{\boldsymbol{\xi}} }}\right] _{\ell,q}}$ can be calculated as 
\begin{equation}	
	\label{eq2.16}	
	\begin{aligned}	
		{ \left[{{\bf J_{\boldsymbol{\xi}} }}\right] _{\ell,q}} & = \frac{2}{{\sigma _n^2}} 2 \Re \left\{ {  \frac{{\partial {\bf y}_0^H}}{\partial {\xi}_\ell} \frac{{\partial {{\bf y}_0}}}{\partial {\xi}_q} } \right\} \\
		&=  \frac{2L}{{\sigma _n^2}}  {\rm tr}  \left( {\bf \Xi}_q {\bf R}_x {\bf \Xi}_\ell^H \right),
	\end{aligned} 
\end{equation}}where $\xi_\ell$ and ${\bf \Xi}_\ell$ denote the $\ell$-th element of $\boldsymbol \xi$ and the $\ell$-th subblock matrix of $\bf \Xi$, $ {\bf R}_x = \mathbb{E}\{{\bf x}(t){\bf x}(t)^H\} = \sum\nolimits_{k = 1}^K {{\bf{w}}{{\bf{w}}^H}}$ denotes the covariance of the transmitted signals. By denoting the ${\bf \Theta} = [\theta_t, \phi_t]^T$ as the AoAs of interest, ${\bf \Psi} = [ \psi_t, \Re(\alpha), \Im(\alpha)]^T$ as the other parameters, the FIM can be partitioned as 
\begin{align}
	\label{eq2.17}
	{\bf J_{\boldsymbol{\xi}}} = \left[ {\begin{array}{*{20}{c}}
			{\bf{J_{\Theta \Theta }}}&{\bf{J_{\Theta \Psi }}}\\
			{{\bf J}_{\bf \Theta \Psi }^H}&{\bf {J_{\Psi \Psi }}}
	\end{array}} \right],
\end{align} 
where ${\bf J_{\boldsymbol{\xi}}} \in \mathbb{C}^{Q\times Q} $, $Q=5$, the ${\bf{J_{\Theta \Theta }}} = {\bf J_{\boldsymbol{\xi}}}(1:2;1:2)$, ${\bf{J_{\Theta \Psi }}} = {\bf J_{\boldsymbol{\xi}}}(1:2;3:5)$, ${\bf {J_{\Psi \Psi }}}= {\bf J_{\boldsymbol{\xi}}}(3:5; 3:5) $. Accordingly, the CRLB matrix for estimating 2D AoAs $\bf \Theta$ is given by \cite{1703855}
\begin{align}
	\label{eq2.18}
	{\rm CRLB}_ {\bf \Theta} = {\left[ {\bf{J_{\Theta \Theta }}} - {{  \bf {J_{\Theta \Psi }} } {\bf J}_{\bf \Psi \Psi }^{ - 1} {\bf J}_{ \bf \Theta \Psi }^H} \right]^{ - 1}}.
\end{align} 
The diagonal elements of the CRLB matrix \eqref{eq2.18} denote the MSE lower bounds of the $\theta_t$ and $\phi_t$. As can be observed, the ${\rm CRLB}_{\bf \Theta}$ depends on both the unknown parameters $\boldsymbol{ \xi}$ and the covariance matrix of transmitted signal ${\bf R}_x$. In practice, the $\boldsymbol{ \xi}$ can be set as the estimation result in the previous frame, which is sufficient for the CRLB design due to the little variation of $\boldsymbol{ \xi}$ between two adjacent coherent time blocks \cite{9652071}. \textcolor{black}{Different from the existing work where only AoAs are taken into consideration \cite{9366836,10050406}, the ${\rm CRLB}_{\bf \Theta}$ in the SSAC system also depends on the AoD ($\psi$) due to the bistatic sensing manner.}

\textcolor{black}{Note that the target sensing can be easily extended to the multiple targets scenario, since the corresponding FIM and CRLB matrix of AoA estimation is in the same form as \eqref{eq2.17} and \eqref{eq2.18}. For simplicity, we focus on the case of one single target.}

\subsection{Problem Formulation}
We aim to maximize the sum rate of VUEs for the SSAC system by optimizing the beamforming matrix subject to the CRLB constraints for the target's AoA estimation. Based on the reconstructed CSI of VUEs with the assistance of sensing information, the problem can be formulated as
\begin{align}
	\label{eq2.19}
	\max _{{\bf{W}}}\quad  &R_{\rm sum}\\
	~{\rm {s.t.}}~\quad   &{\rm Tr} \left( {\rm CRLB}_{\bf \Theta} \right) \le \eta _0, \tag{\ref{eq2.19}a} \label{eq2.19a}\\
	& {\bf{W}} \in \mathcal{W}, \tag{\ref{eq2.19}b} \label{eq2.19b}
\end{align}
where the $\eta_0$ denotes the pre-defined CRLB threshold for the target's 2D-AoA estimation, $\mathcal{W}$ denotes the feasible region of the beamforming matrix, which is different for fully digital array and HAD array as presented in Section IV and V. Note that for VUEs, the communication subspace and the sensing subspace are strongly coupled, which implies that the Fisher information increases with Shannon information when beamforming optimization \cite{10001144}. Therefore, the sensing performance of VUEs no longer serves as a constraint for simplicity, while the CRLB threshold for the weak target is necessary to guarantee the sensing performance.

\section{Beamforming Design for Full-Digital Array}
In this section, we consider the fully digital array setup to draw the important insights between the CRLB optimization and achievable sum rate. Note that the problem \eqref{eq2.19} is highly non-convex due to the fractional form of \eqref{eq2.4} and complicated CRLB matrix \eqref{eq2.19a}. In this section, we propose a PDD-based alternating optimization algorithm to solve the problem \eqref{eq2.19}. Specifically, we first equivalently transform the problem into a more tractable form via the closed-form FP. Then, the problem is decomposed based on the PDD framework, and the corresponding variables are optimized in an alternating manner.

\subsection{Closed-form FP Method}
To address the fractional form, we first equivalently transform the problem \eqref{eq2.19} into a more tractable form. By applying the Lagrangian dual transform \cite{8314727}, the data rate of the $k$-th device can be recast as 
\begin{equation}	
	\label{eq4.A.1}	
	\begin{aligned}	
		\log_2(1+\gamma_k) = &  \max _{\nu _k >0}~  \log_2  \left( {1  + {\nu _k}} \right)  - {\nu _k} \\
		&  + \left( {1 + {\nu _k}} \right) \frac{{{{\left| {{\bf h}_k^H{{\bf w}_k}} \right|}^2}}}{{\sum\limits_{j \in  {\cal K}} {{{\left| {{\bf h}_k^H{{\bf w}_j}} \right|}^2}}  + \sigma _c^2}},
	\end{aligned} 
\end{equation}
where $\nu _k$ is the auxiliary variable. To maximize the value of \textcolor{black}{$k$-th device's data rate}, the variables can be optimized iteratively. With fixed $\bf W$, the optimal $\nu _k$ can be directly calculated by setting $\frac{{\partial {r_k}}}{{\partial {\nu _k}}}=0$, 
\begin{align}
	\label{eq4.A.2}
	\nu_k^{\star} = \frac{{{{\left| {{\bf h}_k^H{{\bf w}_k}} \right|}^2}}}{{\sum\limits_{j \ne k} {{{\left| {{\bf h}_k^H{{\bf w}_j}} \right|}^2}}  + \sigma _c^2}}.
\end{align} 
With fixed $\nu_k$, the date rate maximization problem is equivalent to maximizing the fractional term $\frac{{{{\left| {{\bf h}_k^H{{\bf w}_k}} \right|}^2}}}{{\sum_{j \in  {\cal K}} {{{\left| {{\bf h}_k^H{{\bf w}_j}} \right|}^2}}  + \sigma _c^2}}$. By applying quadratic transform, the problem can be reformulated as
\begin{align}
	\label{eq4.A.3}
	\max _{{{\bf{W}}, \beta_k}}~ & 2\sqrt {1 + {\nu _k}} {\mathop{\Re}\nolimits} \left( {\beta _k^*{\bf h}_k^H{{\bf w}_k}} \right) - {\left| {{\beta _k}} \right|^2}\left( {\sum\limits_{j \in {\cal K}} {{{\left| {{\bf h}_k^H{{\bf w}_j}} \right|}^2}}  + \sigma _c^2} \right)  ,
\end{align}
where $\beta_k$ denotes the auxiliary variable. It is a concave function of $\beta_k$, thus the optimal value can be obtained by setting $\frac{{\partial {f_k}}}{{\partial {\beta _k}}}=0$, resulting in
\begin{align}
	\label{eq4.A.4}
	\beta_k^{\star} = \frac{{\sqrt {1 + {\nu _k}} {\bf{h}}_k^H{{\bf{w}}_k}}}{{\sum\limits_{i \in {\cal K}} {{{\left| {{\bf{h}}_k^H{{\bf{w}}_i}} \right|}^2}}  + \sigma _c^2}}.
\end{align} 

\textcolor{black}{Based on the aforementioned transformation, the problem \eqref{eq2.19} can be formulated as 
\begin{align}
	\label{eq4.A.5}
	\max _{\boldsymbol \nu,\boldsymbol \beta, {\bf{W}}}\quad & r_{\rm sum} \left( \boldsymbol \nu,\boldsymbol \beta, \bf W \right)  \\
	~~~{\rm {s.t.}}~\quad   & {\rm Tr} \left( {\rm CRLB}_{\bf \Theta} \right) \le \eta _0, \tag{\ref{eq4.A.5}a} \label{eq4.A.5a} \\
	&\left\| {\bf W} \right\|_F^2 \le P, \tag{\ref{eq4.A.5}b} \label{eq4.A.5b}
\end{align}}where $r_{\rm sum} \left( \boldsymbol \nu,\boldsymbol \beta, \bf W \right) $ is given as \eqref{eq4.A.6}, $\boldsymbol \nu = [\nu_1,\cdots,\nu_K]^T$ and $\boldsymbol \beta= [\beta_1,\cdots,\beta_K]^T$ denotes the collection of auxiliary variables, ${\cal N} = {\rm{diag}}\left( {2\sqrt {(1 + {\boldsymbol{\nu }})} } \right)$, ${\bf \tilde  H} = \left[{\bf \tilde  h}_1, \cdots, {\bf \tilde  h}_K \right]$, ${\bf \tilde  h}_k = \beta_k {\bf h}_k$,  $P$ denotes the transmit power budget. The problem \eqref{eq4.A.5} can be solved iteratively for each variable, i.e., updating $\boldsymbol \nu$ via \eqref{eq4.A.2}, updating $\boldsymbol \beta$ via \eqref{eq4.A.4}, and updating $\bf W$ via solving \eqref{eq4.A.5}. It can be observed the fraction form is transformed as a quadratic form with respect to the $\bf W$, which is a concave function. Therefore, the only remaining difficulty is the CRLB constraint. 

\begin{figure*}[t]	

\begin{equation}	
	\label{eq4.A.6}	
	\begin{aligned}	
		r_{\rm sum} \left( \boldsymbol \nu,\boldsymbol \beta, \bf W \right) &= \sum\limits_{k \in  {\cal K}} \log_2 \left( {1 + {\nu _k}} \right) - \sum\limits_{k \in  {\cal K}} {\nu _k} + \sum\limits_{k \in  {\cal K}} 2\sqrt {1 + {\nu _k}} {\mathop{\Re}\nolimits} \left( {\beta _k^*{\bf h}_k^H{{\bf w}_k}} \right) - \sum\limits_{k \in  {\cal K}} {\left| {{\beta _k}} \right|^2}\left( {\sum\limits_{j \in {\cal K}} {{{\left| {{\bf h}_k^H{{\bf w}_j}} \right|}^2}}  + \sigma _c^2} \right), \\
		&= {\bf 1}_K^T \left( \log_2 \left( {1 + {\boldsymbol \nu}} \right) -  {\boldsymbol \nu } \right)  + {\mathop{\Re}\nolimits} \left( {tr\left( {{{\mathcal N} {\bf \tilde H}^H {\bf W}}} \right)} \right) - {\left\| {{\bf{\tilde H}}^H {\bf W}} \right\|_F^2} - {\left| \boldsymbol \beta  \right|^2}\sigma _c^2
	\end{aligned} 
\end{equation}

	\hrulefill
\end{figure*}

\subsection{Transformation for CRLB}
By introducing auxiliary semi-positive definite matrix ${\bf{\Omega }} \in \mathbb{C}^{2 \times 2} $, the constraint \eqref{eq2.19a} can be recast as 
\begin{subequations}
 	\begin{align}
 		& {\rm{Tr}}\left( {{{\left[ {{{\bf{J}}_{{\bf{\Theta \Theta }}}} - {{\bf{J}}_{{\bf{\Theta \Psi }}}}{\bf{J}}_{{\bf{\Psi \Psi }}}^{ - 1}{\bf{J}}_{{\bf{\Theta \Psi }}}^H} \right]}^{ - 1}}} \right) \le {\rm{Tr}}\left( {{{\bf{\Omega }}^{ - 1}}} \right), \label{eq4.B.1a} \\
 		& {\rm{Tr}} \left( {{{\bf{\Omega }}^{ - 1}}} \right) \le \eta_0. \label{eq4.B.1b}
 	\end{align}
 \end{subequations}
Note that ${\rm tr}({\bf X}^{-1}) $ is monotonically decreasing on the positive definite matrix space \cite{boyd2004convex}. Therefore, \eqref{eq4.B.1a} is equivalent to the following constraint
\begin{subequations}
	\begin{align}
		& {{\bf{J}}_{{\bf{\Theta \Theta }}}} - {{\bf{J}}_{{\bf{\Theta \Psi }}}}{\bf{J}}_{{\bf{\Psi \Psi }}}^{ - 1}{\bf{J}}_{{\bf{\Theta \Psi }}}^H \succeq \bf \Omega. \label{eq4.B.2a}
	\end{align}
\end{subequations}
Based on the Schur complement condition \cite{boyd2004convex, 10050406}, the CRLB constraint can be reformulated as 
\begin{subequations}
	\begin{align}
		& {\rm Tr}\left( {{{\bf{\Omega }}^{ - 1}}} \right) \le \eta_0,  \label{eq4.2a}\\
		&\left[ {\begin{array}{*{20}{c}}
				{{{\bf{J}}_{{\bf{\Theta \Theta }}}} - {\bf{\Omega }}}&{{{\bf{J}}_{{\bf{\Theta \Psi }}}}}\\
				{{\bf{J}}_{{\bf{\Theta \Psi }}}^H}&{{{\bf{J}}_{{\bf{\Psi \Psi }}}}}
		\end{array}} \right] \succeq 0,  \label{eq4.2b}\\
		&{\bf{\Omega }} \succeq 0.  \label{eq4.2c}
	\end{align}
\end{subequations}

The elements of FIM in \eqref{eq4.2b} are quadratic with respect to the beamforming matrix $\bf {W}$. Thus, the semi-positive definite constraint is still difficult to be solved. To tackle this issue, we first introduce an auxiliary matrix as 
\begin{align}
	\label{eq4.17}
	{\bf F_{\boldsymbol{\xi}}} = \left[ {\begin{array}{*{20}{c}}
			{\bf{F_{\Theta \Theta }}}&{\bf{F_{\Theta \Psi }}}\\
			{{\bf F}_{\bf \Theta \Psi }^H}&{\bf {F_{\Psi \Psi }}}
	\end{array}} \right].
\end{align} 
\textcolor{black}{Then, the problem \eqref{eq2.19} can be recast as 
\begin{align}
	\label{eq4.5}
	\max _{\boldsymbol \nu,\boldsymbol \beta, {\bf{W}, \bf{\Omega}}, {\bf F_{\boldsymbol{\xi}}}}~  & r_{\rm{sum}} \left( \boldsymbol \nu,\boldsymbol \beta, \bf W \right)  \\
	~{\rm {s.t.}}\qquad     & {\rm {Tr}}\left( {{{\bf{\Omega }}^{ - 1}}} \right) \le \eta_0,  \tag{\ref{eq4.5}a}  \label{eq4.5a}\\
				 			&\left[ {\begin{array}{*{20}{c}}
								{{{\bf{F}}_{{\bf{\Theta \Theta }}}} - {\bf{\Omega }}}&{{{\bf{F}}_{{\bf{\Theta \Psi }}}}}\\
								{{\bf{F}}_{{\bf{\Theta \Psi }}}^H}&{{{\bf{F}}_{{\bf{\Psi \Psi }}}}}
							\end{array}} \right] \succeq 0, \tag{\ref{eq4.5}b}  \label{eq4.5b}\\
							&{\bf{\Omega }} \succeq 0,  \tag{\ref{eq4.5}c} \label{eq4.5c}\\
  	   						&{\bf F_{\boldsymbol{\xi}}} = {\bf J_{\boldsymbol{\xi}}}(\bf W),\tag{\ref{eq4.5}d} \label{eq4.5d}\\
  	   						&\eqref{eq4.A.5b}, \notag
\end{align}
}where the ${\bf J_{\boldsymbol{\xi}}}(\bf W)$ is the FIM calculated based on \eqref{eq2.16} and \eqref{eq2.17}. It is observed that if dropping the equality constraint \eqref{eq4.5d}, the problem \eqref{eq4.5} is a quadratic program (QP) for ${\bf{W}}$ and a semi definite program (SDP) for ${\bf \Omega}$. In the next subsection, we address the constraint by introducing an exact penalty term and developing an efficient PDD-based algorithm to obtain a high-quality solution.

\subsection{PDD-based Algorithm for Fully Digital Beamforming Optimization}
\textcolor{black}{Following the PDD framework \cite{9120361}, we define the Augmented Lagrangian (AL) function as follows
\begin{align}
	\label{eq4.6}
	&\max _{\mathcal{X}} ~  r_{\rm{sum}} \left( \boldsymbol \nu,\boldsymbol \beta, \bf W \right)  - \frac{1}{2 \rho_1 }{\left\| {{\bf J_{\boldsymbol{\xi}}}({\bf W}) - {\bf F_{\boldsymbol{\xi}}} + \rho_1 {\bf{Z}}} \right\|_F^2}\\
	&\ ~{\rm {s.t.}}\quad   \eqref{eq4.A.5b},\eqref{eq4.5a}-\eqref{eq4.5c},  \notag
\end{align}
}where $\mathcal{X}=\{\boldsymbol \nu,\boldsymbol \beta, {\bf{W}, \bf{\Omega}}, {\bf F_{\boldsymbol{\xi}}}, {\bf{Z}}\}$ denotes the set of all optimization variables, $\rho_1 > 0$ denotes the penalty factor, ${\bf{Z}}$ denotes Lagrangian dual variable. By decomposing the problem \eqref{eq4.6}, the variables $\bf F_{\boldsymbol{\xi}}$ and ${\bf W}$ can be decoupled. Then, we propose an alternating optimization method to alternately update each variable. \textcolor{black}{The AL problem is solved by alternately updating variables $\boldsymbol \nu$, $\boldsymbol \beta$, $\bf \Omega$, $\bf F_{\boldsymbol{\xi}}$, $\bf W$ until convergence is realized. Then the dual variables or the penalty factor is updated.}

\subsubsection{Update of $\boldsymbol \nu$ and $\boldsymbol \beta$} 
Given the other variables, the optimal auxiliary variables $\boldsymbol \nu^{\star}$ and $\boldsymbol \beta^{\star}$ introduced by FP can be obtained via \eqref{eq4.A.2} and \eqref{eq4.A.4}, respectively.

\subsubsection{Update of $\bf \Omega$ and $\bf F_{\boldsymbol{\xi}}$} 
Fixing other variables, the problem \eqref{eq4.6} can be simplified as
\begin{align}
	\label{eq4.8}
	&\min _{\bf \Omega, \bf F_{\boldsymbol{\xi}}  }\quad  \frac{1}{2 \rho_1 }{\left\| {{\bf J_{\boldsymbol{\xi}}}(\bf W) - {\bf F_{\boldsymbol{\xi}}} + \rho_1 {\bf{Z}}} \right\|_F^2}\\
	&~{\rm {s.t.}}~\quad   \eqref{eq4.5a}-\eqref{eq4.5c}.  \notag
\end{align}
Problem \eqref{eq4.8} is an SDP problem, which can be optimized by using the interior-point method in \cite{ben2001lectures}.

\subsubsection{Update of $\bf W$} By substituting ${\bf J_{\boldsymbol{\xi}}}(\bf W)$ in \eqref{eq2.16} into the penalty term, it is expressed as 
\begin{align}
	\label{eq4.9}
	&{\left\| {{\bf J_{\boldsymbol{\xi}}}(\bf W) - {\bf F_{\boldsymbol{\xi}}} + \rho_1 {\bf{Z}}} \right\|_F^2} = \sum\limits_{\ell =1}^Q \sum\limits_{q=1}^Q \left | {\rm Tr} \left( {\bf{A}}_{\ell,q} {\bf W} {\bf W}^H  \right) -{\bf \tilde F}_{\ell,q} \right|^2,
\end{align}
where ${\bf{A}}_{\ell,q} = \frac{2L}{{\sigma _n^2}}  {\bf \Xi}_\ell^H {\bf \Xi}_q$, ${\bf{\tilde F}} ={\bf F_{\boldsymbol{\xi}}} - \rho_1 {\bf{Z}} $. ${\bf \tilde F}_{\ell,q}$ denotes the element at the $\ell$-th row and the $q$-th column of ${\bf \tilde F}$. By omitting the terms not related to $\bf W$, the AL problem \eqref{eq4.6} is formulated as
\begin{align}
	\label{eq4.10}
	 \max _{{\bf{W}}}\quad &h_w\left({{\bf{W}}}\right) = {\mathop{\Re}\nolimits} \left( {{\rm Tr}\left( {{{\mathcal N} {\bf \tilde H}^H {\bf W}}} \right)} \right) - {\left\| {{\bf{\tilde H}}^H {\bf W}} \right\|_F^2} \notag \\
	& \quad -  \frac{1}{2 \rho_1 }\sum\limits_{\ell =1}^Q \sum\limits_{q=1}^Q \left | {\rm Tr} \left( {\bf{A}}_{\ell,q} {\bf W} {\bf W}^H  \right) -{\bf \tilde F}_{\ell,q} \right|^2 \\
	~{\rm {s.t.}}~~~ &\eqref{eq4.A.5b}. \quad ~   \notag
\end{align}
It is observed that the objective function is concave of $\bf W$ in the feasible region. To solve the problem \eqref{eq4.10}, we resort to the gradient projection algorithm. We first update the beamforming matrix $\bf W$ via the gradient ascent mechanism. The gradient of $h_w\left({{\bf{W}}}\right)$ is calculated as \eqref{eq4.14} on the top of the next page,
\begin{figure*}[t]	
	\begin{align}
		\label{eq4.14}
		\nabla h_w\left( {\bf{W}} \right) = {{{\bf{\tilde H}}}}{{\cal N}^H} - 2{{{\bf{\tilde H}}}}{\bf{\tilde H}}^H - \frac{1}{\rho_1 }\sum\limits_{\ell  = 1}^Q {\sum\limits_{p = 1}^Q {\left( {{{\bf{A}}_{\ell,q}}{\bf{W}}{{\left( {\rm Tr\left( {{{\bf{A}}_{\ell,q}}{\bf{W}}{{\bf{W}}^H}} \right) - {{{\bf{\tilde F}}}_{\ell,q}}} \right)}^H} + \left( {\rm Tr\left( {{{\bf{A}}_{\ell,q}}{\bf{W}}{{\bf{W}}^H}} \right) - {{{\bf{\tilde F}}}_{\ell,q}}} \right){\bf{A}}_{\ell,q}^H{\bf{W}}} \right)} },
	\end{align}
	\hrulefill
\end{figure*}
and $\bf W$ at the $i$-th iteration is updated as 
\begin{align}
	\label{eq4.12}
	{{{\bf{\tilde W}}}^{[i+1]}} = {{\bf{W}}^{[i]}} + \lambda \nabla h_w\left( {{{\bf{W}}^{[i]}}} \right),
\end{align}
where $\lambda$ denotes the step size. To maintain compliance with the transmit power budget, the solution ${{{\bf{\tilde W}}}^{[i]}}$ is then projected onto the feasible region, i.e.,
\begin{align}
	\label{eq4.12a}
	{\bf{W}}^{[i+1]} \gets {\rm{Pro}}{{\rm{j}}_{\cal W}} \left( {{{\bf{\tilde W}}}^{[i+1]}} \right).
\end{align}
The projection operator is given as
\begin{align}
	\label{eq4.13}
	{\rm{Pro}}{{\rm{j}}_{\cal W}} = \left\{ {\begin{array}{*{20}{c}}
			{{\bf{W}},}&{{\rm{if~}} {{\left\| {\bf{W}} \right\|_F}^2} \le P },\\
			{\sqrt P \frac{{\bf{W}}}{{{{\left\| {\bf{W}} \right\|_F}}}},}&{{\rm{otherwise}}{\rm{.}}}
	\end{array}} \right.
\end{align}

The gradient ascent and projection operation are executed alternately until the objective value \eqref{eq4.10} converges.

\subsubsection{Update of $\rho_1$ and ${\bf Z}$}
First, we denote the constraint violation as
\begin{align}
	\label{eq4.15}
	{\mathcal H}^{[n]} = {\left\|  {{\bf J_{\boldsymbol{\xi}}}({\bf W}^{[n]}) - {{\bf F}_{\boldsymbol{\xi}}^{[n]}} } \right\|_\infty },
\end{align}
where $n$ is the index of outer layer iteration. \textcolor{black}{Then, $\rho_1$ and ${\bf{Z}}$ can be updated respectively based on the constraint violation. Specifically, when ${\mathcal H}^{[n]}$ is small, the PDD-based algorithm works the same as the AL method, and the dual variable ${\bf{Z}}$ is updated as
\begin{align}
	\label{eq4.16}
	{\bf{Z}}^{[n]} = {\bf{Z}}^{[n-1]} + \frac{1}{\rho_1} \left( {{\bf J_{\boldsymbol{\xi}}}({\bf W}^{[n]})- {{\bf F}_{\boldsymbol{\xi}}^{[n]}} }  \right).
\end{align}
When ${\mathcal H}^{[n]}$ is large, the penalty factor $\rho_1$ will be decreased to accelerate convergence speed. Both $\rho_1$ and $\bf Z$ will be reduced to zero to satisfy the constraint \eqref{eq4.5d}. }

\subsection{Overall Algorithm and Complexity Analysis}
Based on the algorithms presented above, the overall PDD-based algorithm is presented in Algorithm 1.

\textbf{Initialization:} \textcolor{black}{Before starting up the alternating optimization, the beamforming matrices and auxiliary variables need to be initialized, i.e.,  $\mathcal{X}=\{\boldsymbol \nu,\boldsymbol \beta, {\bf{W}, \bf{\Omega}}, {\bf F_{\boldsymbol{\xi}}}, {\bf{Z}}\}$ and $\rho_1$. Here we present a heuristic initialization scheme with the intuition of radiating the signals towards the direction of the VUEs and the weak target. Specifically, the initial beamforming matrices are set as ${{\bf{W}}^{[0]}} = \sqrt P \frac{{{{\bf{W}}_i}}}{{{{\bf{W}}_i}}}$, in which
\begin{align}
	\label{eq4.16.1}
	{{\bf{W}}_i} = (1-\mu) \left[  {{\bf{w}}_1^{[0]},{\bf{w}}_2^{[0]}, \cdots ,{\bf{w}}_K^{[0]}} \right] + \mu {{\bf{w}}_t}{\bf 1}_K^T,
\end{align}
where ${\bf{w}}_k^{[0]} = {\bf{b}} (\psi_k)$, ${\bf{w}}_t^{[0]} = {\bf{b}} (\psi_t)$, $\mu \in [0,1]$ denotes the weighting factor. Hence, the CRLB constraint can be easily satisfied via one-dimensional searching of $\mu$ from 0 to 1. The auxiliary variables of FP transformation $\boldsymbol \nu^{[0]}$ and $\boldsymbol \beta^{[0]}$ are initialized via \eqref{eq4.A.2} and \eqref{eq4.A.4}, respectively. Besides, we set ${\bf{\Omega}}^{[0]}= \bf 0$, ${\bf F_{\boldsymbol{\xi}}}^{[0]} = \bf 0$, ${\bf{Z} }^{[0]} = \bf 0$. The penalty factor is initialized to balance the weight of the two terms in \eqref{eq4.6}, i.e.,
\begin{align}
	\label{eq4.16.2}
	\rho _1^{[0]} = \frac{{\left\| {{{\bf{J}}_{\boldsymbol{\xi}}}{{({\bf{W}}^{[0]})}} - {\bf{F}}_{\boldsymbol{\xi}}^{[0]} + {\rho _1}{{\bf{Z}}^{[0]}}} \right\|_F^2}}{{2{r_{\rm sum}}\left( {{{\boldsymbol{\nu }}^{[0]}},{{\boldsymbol{\beta }}^{[0]}},{{\bf{W}}^{[0]}}} \right)}}.
\end{align}
}

\textbf{Convergence Analysis:} \textcolor{black}{The proposed PDD-based beamforming design algorithm is a double-loop iterative algorithm, where the inner loop solves the AL subproblem \eqref{eq4.6}, while the outer loop updates the dual variable or the penalty factor. The AL subproblem is solved in an alternating manner. Specifically, the variables $\boldsymbol \nu^{\star}$, $\boldsymbol \beta^{\star}$, $\bf \Omega$, $\bf F_{\boldsymbol{\xi}}$, and $\bf W$ are alternately updated to maximize the objective value of \eqref{eq4.6}. Therefore, the objective value is monotonically non-increasing with the sub-problems alternately solved, and the solution is guaranteed to converge to a stationary point in polynomial time \cite{razaviyayn2013unified}. For the outer loop, the PDD method adaptively switches between the AL and the penalty method. This adaptive strategy is expected to find an appropriate penalty factor, with which the AL method could eventually converge.  For more detailed convergence proof of the PDD framework, we refer readers to \cite{9120361}.}

\textbf{Computational Complexity Analysis:} The auxiliary variables $\boldsymbol \nu,\boldsymbol \beta$, and  ${\bf{Z}}$ are updated by using closed-form expressions, thus computationally efficient. Therefore, the computational complexity of the proposed PDD-based algorithm mainly results from the update of ${\bf F_{\boldsymbol{\xi}}}$, $\bf{\Omega}$, and $\bf W$. The complexity for updating variables ${\bf F_{\boldsymbol{\xi}}}$, and $\bf{\Omega}$ via interior-point methods is ${\mathcal{O}\left(Q^{3.5} \log(1/\epsilon)\right)}$, where $\epsilon$ is the convergence threshold. The complexity complexity of $\bf W$ is ${\mathcal{O}\left(N_i Q^{2} M^2 K\right)}$, which is dominated by calculating the gradient \eqref{eq4.14}. The $N_i$ denotes the iteration number of inner iterations. Therefore, the total complexity is ${\mathcal{O}\left( N_o \left(Q^{3.5} \log(1/\epsilon )+ N_i Q^{2} M^2 K\right)\right)}$, where $N_o$ denotes the number of outer iterations.

\begin{algorithm}[t]
	\caption{Proposed PDD-based Beamforming Design Algorithm for fully digital Array}
	\begin{algorithmic}[1]
		\REQUIRE Feasible $\mathcal{X}^0$, Maximum iteration number, constraint violation threshold $\varepsilon$, $\rho_1>0$, $c \in (0,1)$.

		\STATE Set iteration number $n=1$. 
		\REPEAT		

		\REPEAT	
		\STATE Update $\boldsymbol \nu$ and $\boldsymbol \beta$ via \eqref{eq4.A.2} and \eqref{eq4.A.4}.
		\STATE Update $\bf \Omega$ and $\bf F_{\boldsymbol{\xi}}$ by solving \eqref{eq4.8}.
		\STATE Update $\bf W$ by solving \eqref{eq4.10}.
		\UNTIL {The objective value of AL function is converged or maximum iteration is reached.}
		
		\IF{${\mathcal H}^{[n]} \leq \varepsilon$} 
		\STATE Update the dual variable ${\bf Z}$ via \eqref{eq4.16}.
		\STATE $\rho_1^{[n]} = \rho_1^{[n-1]}$.
		
		\ELSE 
		
		\STATE $\rho_1^{[n]} = c \rho_1^{[n-1]}$, ${\bf Z}^{[n]} = {\bf Z}^{[n-1]}$.
		
		\ENDIF
		
		\STATE Update $n=n+1$.
		
		\UNTIL {The sum rate is converged or maximum iteration is reached.}
		\ENSURE The optimized beamforming matrix $\bf W$.
	\end{algorithmic}
\end{algorithm}

\section{Beamforming Design for HAD Array}
In this section, we consider the HAD array setup. In particular, we propose an effective alternating optimization algorithm based on the proposed PDD-based algorithm for the fully-connected architecture and partially-connected architecture, respectively.

\subsection{Hybrid Beamformer Architecture}
Consider the BS employs the HAD array with $M$ antennas and $N_{rf}$ RF chains, which satisfies $K \le N_{rf} \le M$. The transmitted signal can be written as 
\begin{align}
	\label{eq5.1}
	{\bf x}(t) = \sum\nolimits_{k = 1}^K {{\bf w}_k s_k(t) } = \sum\nolimits_{k = 1}^K {{\bf F}_{a} {\bf d}_k s_k(t) },
\end{align}
where  ${\bf F}_d = \left[{\bf d}_{1},{\bf d}_{2}, \cdots, {\bf d}_{K} \right] \in \mathbb{C}^{N_{rf} \times K}$ denotes the digital beamforming matrix, ${\bf F}_{a} \in \mathbb{C}^{M \times N_{rf}}$ denotes the analog RF beamformer. For the fully-connected array and the partially-connected array, the architecture of ${\bf F}_{a}$ is different.
\subsubsection{Fully-Connected Architecture} Each RF chain is connected to all the antennas. The analog beamformer is an $M \times N_{rf}$  dimensional matrix with unit-modulus elements, with the feasible region expressed as ${{\cal F}_f} = \left\{ {{\bf{F}} \in {^{M \times {N_{rf}}}}|\left| {{{\bf{F}}_{i,j}}} \right| = 1,\forall i,j} \right\}$.

\subsubsection{Partially-Connected Architecture} To further reduce the hardware cost and complexity, each RF chain in the partially connected array is only connected with $M/N_{rf}$ antennas. The feasible region $\mathcal{F}_p$ of the analog beamformer can be characterized by the following block diagonal matrix,
\begin{align}
	\label{eq5.1a}
	{{\bf{F}}_{a}} = \left[ {\begin{array}{*{20}{c}}
			{{{\bf{p}}_1}}&0& \cdots &0\\
			0&{{{\bf{p}}_2}}&{}&0\\
			\vdots &{}& \ddots & \vdots \\
			0&0& \cdots &{{{\bf{p}}_{{N_{a}}}}}
	\end{array}} \right],
\end{align}
where ${\bf p}_i$ is an $N_{p}=M/N_{rf}$ dimension vector with unit-modulus elements, i.e., $|\left({\bf p}_{i}\right)_{j}| = 1$.

Based on the PDD framework in Algorithm 1, the key to optimizing the hybrid beamformer is to solve the problem \eqref{eq4.10} with the digital beamforming matrix $\bf W$ replaced with ${\bf F}_{a} {\bf F}_d$, which is reformulated as
\begin{align}
	\label{eq5.4}  \max _{{\bf{F}}_{a}, {\bf F}_d}\quad  & \Re \left( {{\rm Tr}\left( {{\cal N}{\bf{\tilde H}}^H{{\bf{F}}_{a}}{{\bf{F}}_d}} \right)} \right) - {{\left\| {{\bf{\tilde H}}^H{{\bf{F}}_{a}}{{\bf{F}}_d}} \right\|_F}^2} \notag \\
	              & -  \frac{1}{2 \rho_1 }\sum\limits_{\ell =1}^Q \sum\limits_{q=1}^Q \left | {\rm Tr} \left( {\bf{A}}_{\ell,q} {\bf{F}}_{a} {\bf F}_d {\bf F}_d^H {\bf{F}}_{a}^H  \right) -{\bf \tilde F}_{\ell,q} \right|^2,                                 \\
	              	~{\rm {s.t.}}~\quad   & \left\|{\bf F}_{a} {\bf F}_d \right\|_F^2 \le P, \tag{\ref{eq5.4}a} \label{eq5.4a} \\
	              	& { {{\bf{F}}_{a}} } \in \mathcal{F}_{a}, \tag{\ref{eq5.4}b} \label{eq5.4b}
\end{align}
where ${{\mathcal{F}}_{a}} \in \{{{\mathcal{F}}_{f}}, {{\mathcal{F}}_{p}}   \}$ denotes the feasible region of the analog beamformer.
We propose an alternating optimization algorithm to jointly design the hybrid beamformer. The optimization of the digital beamformer is valid for both fully-connected and partially-connected HAD arrays. As for the analog beamformer, two algorithms are designed for the two architectures, respectively.

\subsection{Optimization of Digital Beamformer}
With fixed ${\bf F}_{a}$, the format of problem \eqref{eq5.4} is similar to \eqref{eq4.10}, which can be solved by gradient projection algorithm. Let us first compute the gradient of \eqref{eq4.10} as $\nabla h_d\left( {\bf F}_d\right)$ in \eqref{eq5.5}, where ${{{\bf{\tilde A}}}_{\ell,q}} = {\bf{F}}_{a}^H{{\bf{A}}_{\ell,q}}{{\bf{F}}_{a}}{\left( {{\rm{Tr}}\left( {{\bf{F}}_{a}^H{{\bf{A}}_{\ell,q}}{{\bf{F}}_{a}}{{\bf{F}}_d}{\bf{F}}_d^H} \right){\rm{ - }}{{{\bf{\tilde F}}}_{\ell ,{\rm{p}}}}} \right)^H} $.

\begin{figure*}[t]	
	
\begin{align}
	\label{eq5.5}
	\nabla h_d\left( {\bf F}_d\right) = {\bf{F}}_{a}^H{{\bf{\tilde H}}}{{\cal{N}}^H} - 2{\bf{F}}_{a}^H{{\bf{\tilde H}}}{\bf{\tilde H}}^H{{\bf{F}}_{a}}{{\bf{F}}_d} - \frac{1}{{{\rho _1}}}\sum\limits_{\ell  = 1}^Q {\sum\limits_{p = 1}^Q {\left( {\left( {{{{\bf{\tilde A}}}_{\ell,q}} + {\bf{\tilde A}}_{\ell,q}^H} \right){{\bf F}_d}} \right)} },
\end{align}
	
	\hrulefill
\end{figure*}

Then, we update the digital beamforming matrix until convergence via 
\begin{align}
	\label{eq5.6}
	{\bf{F}}_d^{[n+1]} \gets {\rm{Pro}}{{\rm{j}}_{{\cal F}_d}} \left( {\bf{F}}_d^{[n]} + \lambda \nabla h\left( {\bf{F}}_d^{n} \right) \right),
\end{align}
where ${{\cal F}_d}$ denotes the feasible region of the digital beamforming matrix ${\bf F}_d$. The projection operator is given as
\begin{align}
	\label{eq5.7}
	{\rm{Pro}}{{\rm{j}}_{{\cal F}_d}} = \left\{ {\begin{array}{*{20}{c}}
			{{\bf{F}}_d,}&{{\rm{if~}} {{\left\| {\bf{F}}_{a} {\bf{F}}_d \right\|_F}^2} \le P },\\
			{\sqrt P \frac{{\bf{F}}_d}{{{{\left\| {\bf{F}}_{a} {\bf{F}}_d  \right\|_F}}}},}&{{\rm{otherwise}}{\rm{.}}}
	\end{array}} \right.
\end{align}

\begin{figure*}[t]	
	\vspace{-0.5cm}
	\begin{align}
		\label{eq5.10}
		& \nabla h_f\left( {{{\bf{F}}_{a}}} \right) = {{{\bf{\tilde H}}}}{{\cal{N}}^H}{\bf{F}}_d^H - 2{\bf{\tilde H}}^H{{{\bf{\tilde H}}}}{{\bf{F}}_{a}}{{\bf{F}}_d}{\bf{F}}_d^H - \frac{1}{{{\rho _1}}}\sum\limits_{\ell  = 1}^Q {\sum\limits_{p = 1}^Q {\left( {\left( {{{{\bf{\hat A}}}_{\ell,q}} + {\bf{\hat A}}_{\ell,q}^H} \right){{\bf{F}}_{a}}{{\bf{F}}_d}{\bf{F}}_d^H} \right)} }  - \frac{1}{{{\rho _2}}}\nabla \Phi \left( {{{\bf{F}}_{a}}} \right), \tag{50}
	\end{align}
	\hrulefill
	\begin{align}
		\label{eq5.11}
		& \nabla \Phi \left( {{{\bf{F}}_{a}}} \right) = \left\{ {\begin{array}{*{20}{c}}
				{4\left( {{{\left\| {{{\bf{F}}_{a}}{{\bf{F}}_d}} \right\|_F}^2} - P} \right){{\bf{F}}_{a}}{{\bf{F}}_d}{\bf{F}}_d^H}&{{\rm{if }}{{\left\| {{{\bf{F}}_{a}}{{\bf{F}}_d}} \right\|_F}^2} > P},\\
				0&{{\rm{otherwise}}},
		\end{array}} \right.\tag{51}
	\end{align}
	\hrulefill
\end{figure*}

\subsection{Optimization of Fully-Connected Analog Beamformer}
\textcolor{black}{
As can be seen, with fixed ${\bf F}_d$, the optimization of ${\bf F}_{a}$ is hindered by the power constraint and the unit-modulus constraint. To address this issue, we incorporate the power constraint into the objective function as a penalty term. Then, problem \eqref{eq5.4} is reformulated as 
\begin{subequations}
	\begin{align}
	& {\max _{{{\bf{F}}_{a}} \in \mathcal F_f } } ~ h_f\left( {{{\bf{F}}_{a}}} \right) = \Re \left( {{\rm Tr}\left( {{\cal N}{\bf{\tilde H}}^H{{\bf{F}}_{a}}{{\bf{F}}_d}} \right)} \right) - {{\left\| {{\bf{\tilde H}}^H{{\bf{F}}_{a}}{{\bf{F}}_d}} \right\|_F}^2} \notag\\
	& - \frac{1}{{2\rho_1 }}\sum\limits_{\ell  = 1}^Q {\sum\limits_{p = 1}^Q {{{\left| {{\rm{Tr}}\left( {{{\bf{A}}_{\ell,q}}{{\bf{F}}_{a}}{{\bf{F}}_d}{\bf{F}}_d^H{\bf{F}}_{a}^H} \right) - {{{\bf{\tilde F}}}_{\ell,q}}} \right|}^2}} }  - \frac{1}{{{\rho_2}}} \Phi({{\bf{F}}_{a}}) 	\label{eq5.8}\\
	&~{\rm {s.t.}}~ |\left[{\bf{F}}_{a} \right]_{i,j} |= 1, \forall i,j,   \label{eq5.8a}
	\end{align}
\end{subequations}
}where $\Phi ({{\bf{F}}_{a}}) = {\left( {{{\left[ {{{\left\| {{{\bf{F}}_{a}}{{\bf{F}}_d}} \right\|_F}^2} - P} \right]}^ + }} \right)^2}$  denote the square penalty term, ${\left[ \cdot \right]^ + }$ denotes ${\rm{max}}\left\{ {\cdot,0} \right\}$, $\rho_2$ denotes the weighting factor of the penalty term, which is decreasing with iterations. The feasible region $\mathcal{F}_f$ is indeed a complex circle manifold. Therefore, the problem \eqref{eq5.8} is reformulated as an unconstrained problem in the complex circle manifold, where the manifold optimization can be employed to solve it \cite{Absil2009Optimization}.

The \textit{tangent space} at a point $\bf X$ of the manifold $\mathcal{F}_f$ is defined as the set of all the tangent vectors at $\bf X$, which is expressed as
\begin{align}
	\label{eq5.9}
	& {T_{\bf{X}}}{{\cal F}_f} = \left\{ {{\bf{V}} \in {^{M \times {N_{rf}}}}|\Re \left\{ {{\bf{X}} \odot {\bf{V}}} \right\} = {{\bf{0}}_{M \times {N_{rf}}}}} \right\},
\end{align}
where ${\bf{V}}$ denotes a tangent vector at $\bf X$, ${{\bf{0}}_{M \times {N_{rf}}}}$ denotes the $M \times {N_{rf}}$ dimensional matrix with all zero elements.
The steepest ascent direction of the objective function $h_f\left( {{{\bf{F}}_{a}}} \right)$ in tangent space ${T_{\bf{X}}}{{\cal F}_f}$ is defined as \textit{Riemannian gradient}. To calculate the Riemannian gradient, we first compute the Euclidean gradient of $h_f\left( {{{\bf{F}}_{a}}} \right)$, which is given as $ \nabla h_f\left( {{{\bf{F}}_{a}}} \right)$ in \eqref{eq5.10} and \eqref{eq5.11}\addtocounter{equation}{2}, where ${{{\bf{\hat A}}}_{\ell,q}} = {{\bf{A}}_{\ell,q}}{\left( {{\rm{Tr}}\left( {{{\bf{A}}_{\ell,q}}{{\bf{F}}_{a}}{{\bf{F}}_d}{\bf{F}}_d^H{\bf{F}}_{a}^H} \right) - {{{\bf{\tilde F}}}_{\ell,q}}} \right)^H}$.

Accordingly, the Riemannian gradient ${\rm grad } h_f$ is calculated by projecting Euclidean gradients onto the tangent space, i.e.,
\begin{align}
	\label{eq5.12}
	& {\rm grad}h_f\left( {{{\bf{F}}_{a}}} \right) = \nabla h_f\left( {{{\bf{F}}_{a}}} \right) - \Re \left( {\nabla h_f\left( {{{\bf{F}}_{a}}} \right) \odot {\bf{F}}_{a}^ * } \right) \odot {{\bf{F}}_{a}}.
\end{align}

To find the maximum of $h_f$, the ascent direction is chosen as the Riemannian gradient. Given ${\bf F}_{a} \in \mathcal F_p$ and retraction $\mathcal R$ on $\mathcal F_p$, the variable $ {\bf F}_{a}$ can be updated iteratively as 
\begin{align}
	\label{eq5.13}
	& {\bf{F}}_{a}^{[n+1]} \gets {{\cal R}_{{\bf{F}}_{a}^{[n]}}}\left( {\lambda {\rm{grad}} h_f\left( {{\bf{F}}_{a}^{[n]}} \right)} \right),
\end{align}
where $n$ denotes the iteration index, ${{\cal R}_{{\bf{F}}_{a}^{[n]}}}$ is utilized to map the updated variable in the tangent space ${T_{\bf{X}}}{{\cal F}_f}$ onto the complex circle manifold ${\cal F}_f$, which is chosen as
\begin{align}
	\label{eq5.14}
	& {R_{{\bf{F}}_{a}^{[n]}}}\left( {\lambda {\rm{grad}}h_f} \right) = \frac{{{{\left( {{\bf{F}}_{a}^{[n]} + \lambda {\rm{grad}}h_f} \right)}_{i,j}}}}{{\left| {{{\left( {{\bf{F}}_{a}^{[n]} + \lambda {\rm{grad}}h_f} \right)}_{i,j}}} \right|}}.
\end{align}

\subsection{Optimization of Partially-Connected Analog Beamformer}
For the partially-connected beamformer, the feasible region is not a complex circle manifold. Therefore, the aforementioned manifold optimization algorithm cannot be directly applied. Considering the sparsity of the special architecture, we develop a more computationally efficient algorithm. Specifically, we first reformulate the problem \eqref{eq5.4} as
\begin{align}
	\label{eq5.15} & \max _{{\bf{F}}_{a}}\quad h_{p0}\left({\bf{F}}_{a} \right)  \\
	& ~{\rm {s.t.}}~ { {{\bf{F}}_{a}} } \in \mathcal{F}_{p}, \tag{\ref{eq5.15}a} \label{eq5.15a}
\end{align}
where $h_{p0}\left({\bf{F}}_{a} \right)$ is given as \eqref{eq5.16}. 
\begin{figure*}[t]	
	\vspace{-0.5cm}
	\begin{align}
		\label{eq5.16}
		&  h_{p0}\left( {{{\bf{F}}_{a}}} \right) = \sum\limits_{k \in {\cal K}} {\Re \left( {{\rm Tr}\left( {2\sqrt {1 + {\nu _k}} {{{\bf{\tilde h}}}_k^H}{{\bf{F}}_{a}}{{\bf{d}}_{k}}} \right)} \right)}  - \sum\limits_{k \in {\cal K}} {{{\left\| {{{{\bf{\tilde H}}}^H}{{\bf{F}}_{a}}{{\bf{d}}_{k}}} \right\|_F}^2}}  - \frac{1}{{2{\rho _1}}}\sum\limits_{\ell  = 1}^Q {\sum\limits_{p = 1}^Q {{{\left| {\sum\limits_{k \in {\cal K}} {{\rm{Tr}}\left( {{{\bf{A}}_{\ell,q}}{{\bf{F}}_{a}}{{\bf{d}}_{k}}{\bf{d}}_{k}^H{\bf{F}}_{a}^H} \right)}  - {{{\bf{\tilde F}}}_{\ell,q}}} \right|}^2}} } ,  
	\end{align}
	\hrulefill
\end{figure*}
Note that the transmit power can be recast as 
\begin{align}
	\label{eq5.17} 
	\left\|{\bf F}_{a} {\bf F}_d \right\|_F^2 = \frac{M}{N_{rf}}\left\| {\bf F}_d \right\|_F^2 = N_{p} \left\| {\bf F}_d \right\|_F^2,
\end{align}
which is irrelevant to ${\bf F}_{a}$. The power constraint \eqref{eq5.4a} can be dropped when optimizing the partially-connected analog beamformer. By applying the matrix transformation, the transmitted vector ${{\bf{F}}_{a}}{{\bf{d}}_{k}}$ for user $k$ can be written as
\begin{equation}
	\label{eq5.18}
	\begin{aligned}
		{{\bf{F}}_{a}}{{\bf{d}}_k} &= \left[ {\begin{array}{*{20}{c}}
				{{{\bf{p}}_1}}&0& \cdots &0\\
				0&{{{\bf{p}}_2}}&{}&0\\
				\vdots &{}& \ddots & \vdots \\
				0&0& \cdots &{{{\bf{p}}_{{N_{rf}}}}}
		\end{array}} \right]\left[ {\begin{array}{*{20}{c}}
				{{d_{k,1}}}\\
				{{d_{k,2}}}\\
				\vdots \\
				{{d_{k,{N_{rf}}}}}
		\end{array}} \right]\\
		& = \left[ {\begin{array}{*{20}{c}}
				{{d_{k,1}}{{\bf{p}}_1}}\\
				{{d_{k,2}}{{\bf{p}}_2}}\\
				\vdots \\
				{{d_{k,{N_{rf}}}}{{\bf{p}}_{{N_{rf}}}}}
		\end{array}} \right] = {{\bf{D}}_k}{\bf{p}},
	\end{aligned} 
\end{equation}
where ${{\bf{D}}_k} = {\rm diag}\left( {{{\bf{d}}_k} \otimes {{\bf{1}}_{{N_p}}}} \right) \in \mathbb{C}^{M_t \times M_t}$, ${\bf{p}} = {\left[ {{\bf{p}}_1^H,{\bf{p}}_2^H, \cdots, {\bf{p}}_{{N_{rf}}}^H} \right]^H} \in \mathbb{C}^{M_t \times 1} $. By substituting ${{\bf{F}}_{a}}{{\bf{d}}_{k}}$ into \eqref{eq5.16}, the problem \eqref{eq5.15} can be recast as 
\begin{align}
	\label{eq5.19} & \max _{{\bf{p}}}\quad h_p\left({\bf{p}} \right)  \\
	& ~{\rm {s.t.}}~\quad \left| {{p_j}} \right| = 1,\forall j = 1, \cdots ,M, \tag{\ref{eq5.19}a} \label{eq5.19a}
\end{align}
where $h_p\left({\bf{p}} \right)$ is given as \eqref{eq5.20}. 
\begin{figure*}[t]	
	\vspace{-0.5cm}
	\begin{align}
		\label{eq5.20}
		&  h_p\left( {\bf{p}} \right) = \sum\limits_{k \in K} {\Re \left( {{\rm Tr}\left( {2\sqrt {1 + {\nu _k}} {\bf{\tilde h}}_k^H{{\bf{D}}_k}{\bf{p}}} \right)} \right)}  - \sum\limits_{k \in K} {{{\left\| {{{{\bf{\tilde H}}}^H}{{\bf{D}}_k}{\bf{p}}} \right\|_F}^2}}  - \frac{1}{{2{\rho _1}}}\sum\limits_{\ell  = 1}^Q {\sum\limits_{p = 1}^Q {{{\left| {\sum\limits_{k \in K} {{\rm{Tr}}\left( {{{\bf{A}}_{\ell,q}}{{\bf{D}}_k}{\bf{p}}{{\bf{p}}^H}{\bf{D}}_k^H} \right)}  - {{{\bf{\tilde F}}}_{\ell,q}}} \right|}^2}} } ,  
	\end{align}
	\hrulefill
	\begin{align}
		\label{eq5.23}
		& \nabla h_p\left( {\bf{p}} \right) = \sum\limits_{k \in K} {2\sqrt {1 + {\nu _k}} {\bf{D}}_k^H{{{\bf{\tilde h}}}_k}}  - 2\sum\limits_{k \in K} {\left( {{\bf{D}}_k^H{\bf{\tilde H}}{{{\bf{\tilde H}}}^H}{{\bf{D}}_k}{\bf{p}}} \right)}  - \frac{1}{{{\rho _1}}}\sum\limits_{\ell  = 1}^Q {\sum\limits_{p = 1}^Q {\left( {\left( {{{{\bf{\bar A}}}_{\ell,q}} + {\bf{\bar A}}_{\ell,q}^H} \right){\bf{p}}} \right)} },  
	\end{align}
	\hrulefill
\end{figure*}

We then apply the manifold optimization technique to solve this problem. The Euclidean gradient is calculated as \eqref{eq5.23}, where 
\begin{align}
	\label{eq5.23a}
	 &{{{\bf{\bar A}}}_{\ell,q}} = \notag \\ 
	 &\left( {\sum\limits_{k \in {\cal K}} {{\bf{D}}_k^H{{\bf{A}}_{\ell,q}}{{\bf{D}}_k}} } \right){\left( {\sum\limits_{k \in K} {{\rm{Tr}}\left( {{{\bf{A}}_{\ell,q}}{{\bf{D}}_k}{\bf{p}}{{\bf{p}}^H}{\bf{D}}_k^H} \right)}  - {{{\bf{\tilde F}}}_{\ell,q}}} \right)^H}.
\end{align}
Accordingly, the Riemannian gradient is obtained by 
\begin{align}
	\label{eq5.21}
	& {\rm grad} h_p\left( {{{\bf{p}}}} \right) = \nabla h_p\left( {{{\bf{p}}}} \right) - \Re \left( {\nabla h_p\left( {{{\bf{p}}}} \right) \odot {\bf{p}}^ * } \right) \odot {{\bf{p}}},
\end{align}
Therefore, we can update the beamforming matrix by \textcolor{black}{
\begin{align}
	\label{eq5.22}
	& {\bf{p}}^{{[n+1]}} \gets {{\cal R}_{{\bf{p}}^{[n]}}}\left( {\lambda {\rm{grad}}h_p\left( {{\bf{p}}^{[n]}} \right)} \right).
\end{align}
}

\subsection{Overall Algorithm and Complexity Analysis}
Based on the aforementioned algorithm, the HAD beamformer can be designed via alternating optimizing digital beamformer ${\bf F}_d$ and analog beamformer ${\bf F}_{a}$, which is presented in Algorithm 2. \textcolor{black}{The initialization of the variables and the penalty factor is the same as that of Algorithm 1. The initial digital beamformer and analog beamformer are obtained by decomposing the initial fully digital beamforming matrices via the method in \cite{6966076}.} The objective function \eqref{eq5.4} is non-increasing over iterations. Therefore, Algorithm 2 is guaranteed to converge. The overall sum rate maximization problem with CRLB constraint can be achieved based on the PDD framework, where the step of updating fully digital beamformer $\bf W$ in Algorithm 1 should be replaced by Algorithm 2.

The computational complexity of Algorithm 2 comes from updates for the digital beamformer and analog beamformer, which are dominated by calculating the gradient in \eqref{eq5.5}, \eqref{eq5.10}, or \eqref{eq5.23}. The computational complexity of optimizing digital beamformer is $\mathcal{O}\left(N_{i,d}(N_{rf}^2K+N_{rf}^3)\right)$, where $N_{i,d}$ denote the number of iterations. The computational complexity of analog beamformer is $\mathcal{O}\left(N_{i,f} \left(Q^2 (M^2 N_{rf} + MN_{rf}^3)\right) \right)$ for fully-connected HAD arrays and $\mathcal{O}\left(N_{i,p}(Q^2 M^2)\right)$ for partially-connected HAD arrays,  with $N_{i,f}$ and $N_{i,p}$ being the number of iterations, respectively. As can be observed, the complexity of the design for partially-connected HAD arrays is reduced due to the inherent sparsity.

\begin{algorithm}[t]
	\caption{Alternating algorithm for solving problem \eqref{eq5.4}.}
	\begin{algorithmic}[1]
		\REQUIRE Feasible ${\bf F}_d^{[0]}$, ${\bf F}_{a}^{[0]}$, Maximum iteration number, convergence threshold $\varepsilon$.

		\REPEAT		
		\STATE Set iteration number $n=1$.
		\REPEAT
		\STATE Compute $\nabla h_d$ via \eqref{eq5.5}.
		\STATE Update the ${\bf F}_d^{[n+1]}$ via \eqref{eq5.6}, $n=n+1$.
		\UNTIL The value of objective function $h\left( {\bf F}_d\right)$ is converged or maximum iteration reached.
		
		\STATE Set iteration number $n=1$.
		\IF{Full-connected architecture}
		
		\REPEAT
		\STATE Compute Euclidean gradients $\nabla h_f$ via \eqref{eq5.10}.
		
		\STATE Compute Riemannian gradient $ {\rm grad}~ h_f $ via \eqref{eq5.12}.
		
		\STATE Update ${{{\bf{F}}_{a}^{[n+1]}}}$ via \eqref{eq5.13}, $n=n+1$.
		
		\UNTIL ${\rm grad}~ h_f \le \varepsilon$ or maximum iteration reached.
		
		\STATE Update $\rho_2^{[n+1]} = c \rho_2^{[n]}$.
		
		\ELSIF {Partially-connected architecture} 
		
		\REPEAT
				
		\STATE Compute Euclidean gradient $\nabla h_p$ via \eqref{eq5.23}.
		
		\STATE Compute Riemannian gradient $ {\rm grad}~ h_p$ via \eqref{eq5.21}.
		
		\STATE Update ${{{\bf{p}}^{[n+1]}}}$ via \eqref{eq5.22}, $n=n+1$.
		
		\UNTIL ${\rm grad}~ h_p \le \varepsilon$ or maximum iteration reached.
			
		\ENDIF 
		
		\UNTIL {The objective function \eqref{eq5.4} is converged or maximum iteration is reached.}
		\ENSURE The optimized HAD beamforming matrix ${\bf F}_d$, ${\bf F}_{a}$.
	\end{algorithmic}
\end{algorithm}

\section{Simulation Results}
\begin{figure}[t]
	\centering
	\includegraphics[width=0.8\linewidth]{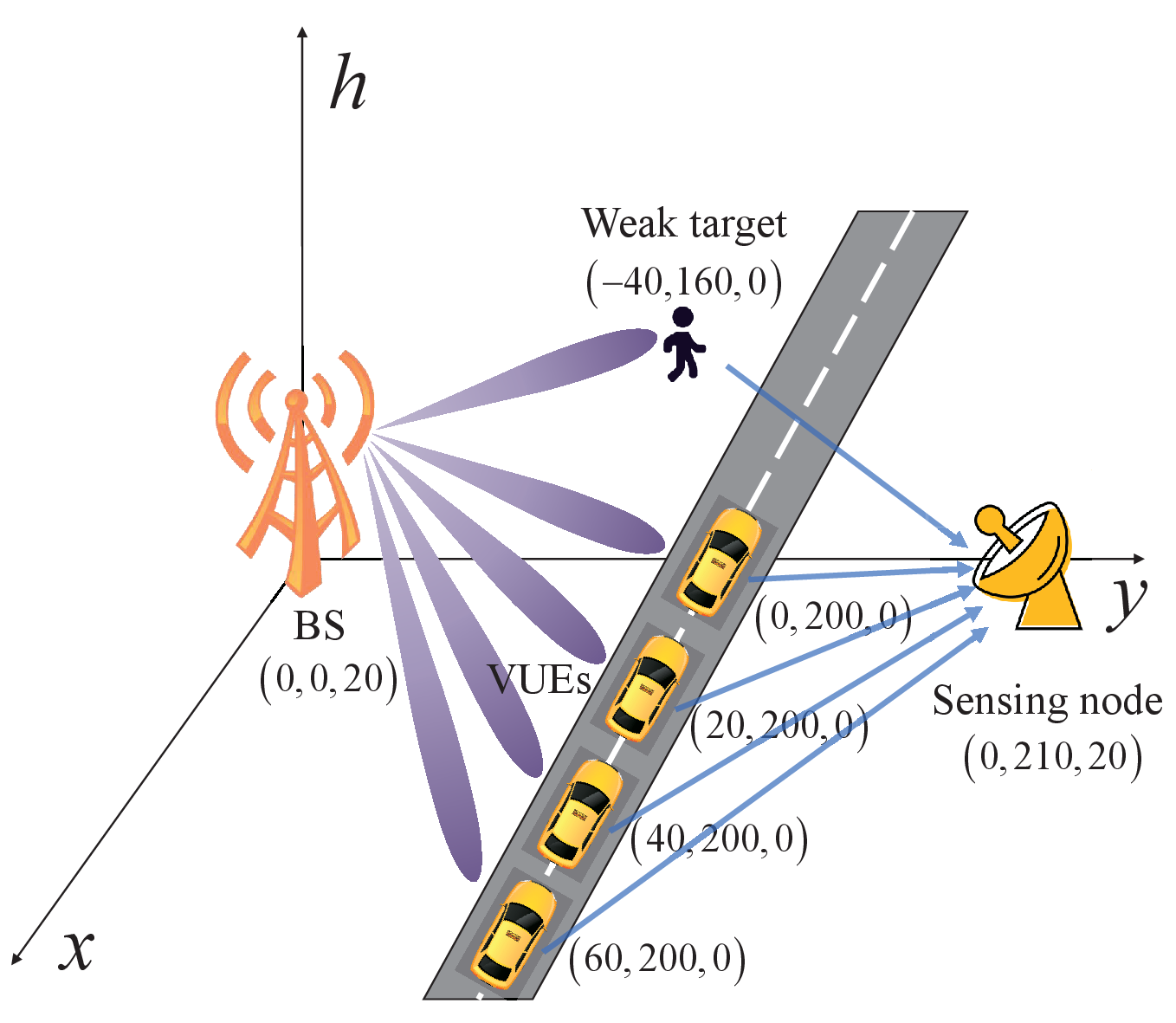}
	\caption{An illustration of the simulation setup.}
	\label{model2}
\end{figure}

In this section, we provide numerical results to verify the effectiveness of the proposed beamforming design algorithms. In the considered SSAC system, a BS and a passive sensing node are located at the two opposite sides of the road to serve $K=4$ single-antenna VUEs and track a weak target as illustrated in \figurename{ \ref{model2}}. The RCS of VUEs and the weak target are set as $\zeta_k = 20$ dBsm and $\zeta_t = 0$ dBsm, respectively. We consider three types of BS, which are equipped with a fully digital array, a fully-connected HAD array, and a partially-connected HAD array, respectively. Unless otherwise stated, the number of BS transmit antennas and the number of receive antennas of the sensing node are set as $M=N=64$. For the BS equipped with an HAD array, we consider the number of RF chains to be equal to that of data streams $N_s$, i.e., $N_{rf} = N_s=K =4$. The noise power at the receivers is set as $\sigma_c^{2}=\sigma_r^{2}= -80$ dBm. The power budget is set as $p = 30$ dBm. \textcolor{black}{The path loss exponent and channel gain at the reference distance of $1$ m are set as $\alpha_0=2$ and $\rho_0 = -50$ dB, respectively. Following \cite{10000676}, the path gain of the NLoS channels is attenuated by $5-10$ dB, i.e., ${\left| {{\alpha _{n,i}}} \right|^2}$ is uniformly distributed between $-10$ dB and $-5$ dB. The number of NLoS paths is $n_{LK}=3$.} We set the CRLB threshold  $\eta=-40 $ dB, the number of snapshots for sensing $L=1000$. \textcolor{black}{All simulation results are averaged over 100 Monte-Carlo experiments.}

\subsection{Sensing Performance Evaluation}
\begin{figure}[t]
	\centering
	\includegraphics[width=0.8\linewidth]{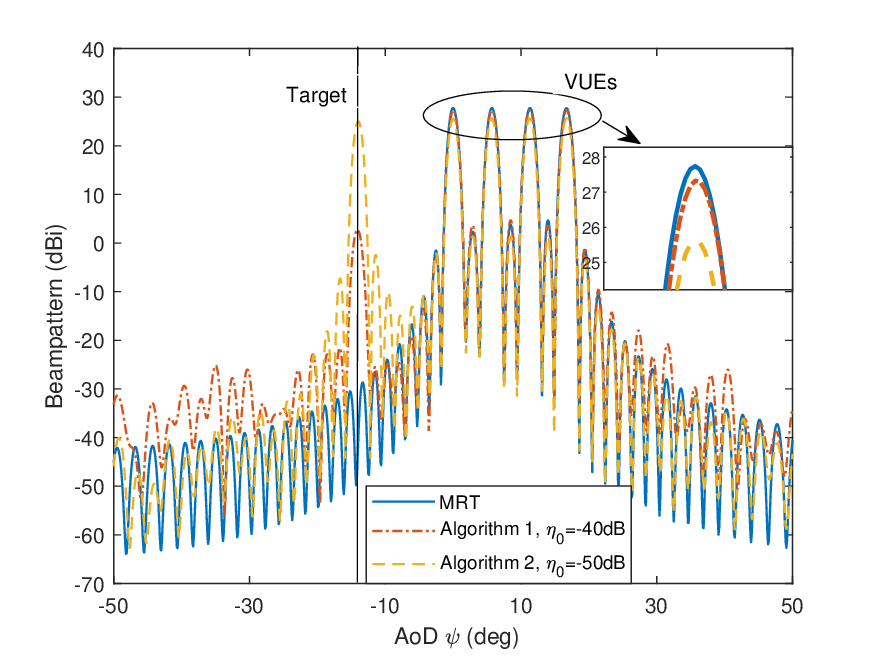}
	\caption{Beampattern comparison with different approaches.}
	\label{beampattern}
\end{figure}

\begin{figure}[t]
	\centering
	\begin{subfigure}[]
		{\centering
			\includegraphics[width=0.45\linewidth]{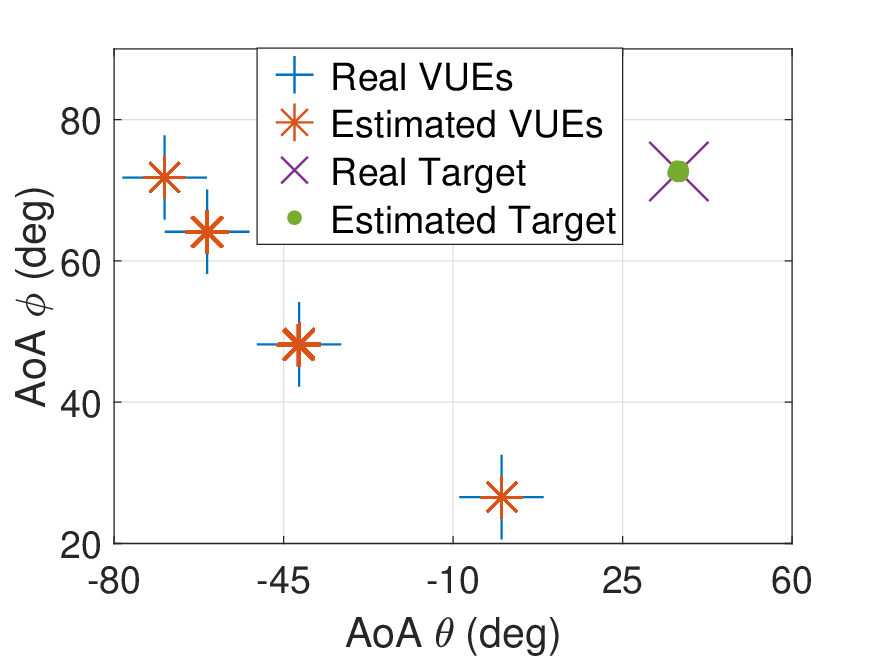}}
	\end{subfigure}
	\begin{subfigure}[]
		{\centering
			\includegraphics[width=0.45\linewidth]{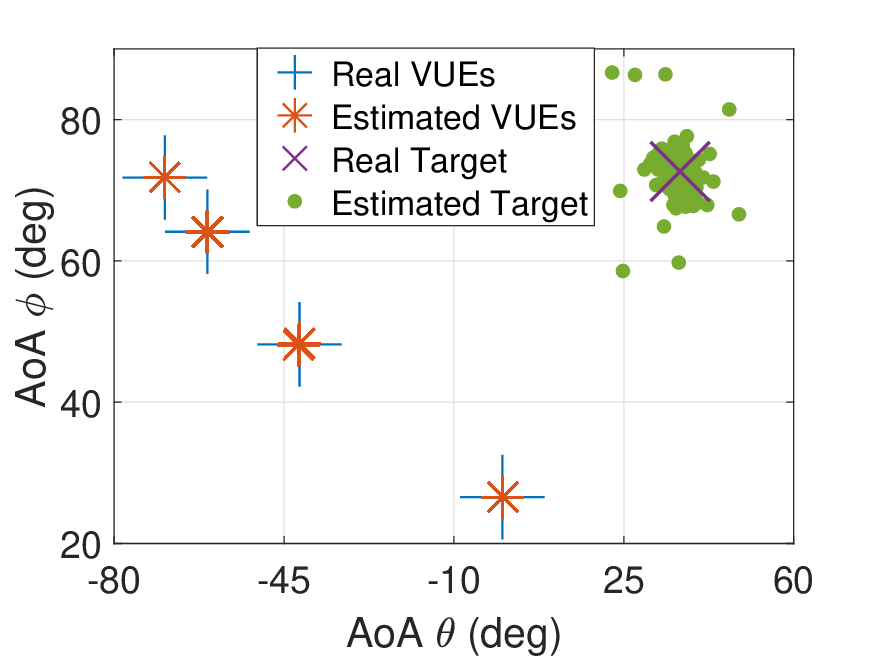}}
	\end{subfigure}
	\caption{Angle estimation results. (a) Beamforming design via Algorithm 1 with CRLB constraint $\eta_0 = -40$ dB; (b) Beamforming design via MRT.}
	\label{DOA}
\end{figure}

In this subsection, we evaluate the sensing performance to verify the effectiveness of the proposed SSAC framework and beamforming design, with the fully digital one taken as an example. In \figurename{ \ref{beampattern}}, we present the beampatterns of our proposed algorithm and the conventional maximum-ratio transmission (MRT) algorithm \cite{765552} in fully digital arrays. It can be observed that the power radiated towards the directions of VUEs and the targets in our proposed algorithm. Moreover, the beam is more concentrated towards the direction of the target with a smaller CRLB threshold. As for the MRT scheme, the beam is directed to the VUEs, while the sensing target is ignored. The angle estimation results for the VUEs and the target are shown in \figurename{ \ref{DOA}}, where the AoAs of VUEs are estimated via the 2D-MUSIC algorithm, and the AoAs of the weak target are obtained by the RD-MLE algorithm with spatial filtering \cite{van2002optimum}.  As shown in \figurename{ \ref{DOA} (a)}, both the VUEs and the target can be located by the beamformer designed with Algorithm 1. As mentioned above, the communication subspace and the sensing subspace are strongly coupled. Even if the sensing performance of VUEs is not taken into consideration in beamforming design, the VUEs can still be accurately located. \figurename{ \ref{DOA} (b)} depicts the sensing results with the MRT algorithm. Without effective constraint on AoA estimation CRLB, the echo signals from the target are too weak, resulting in blurry estimation.

\begin{figure}[t]
	\centering
	\includegraphics[width=0.8\linewidth]{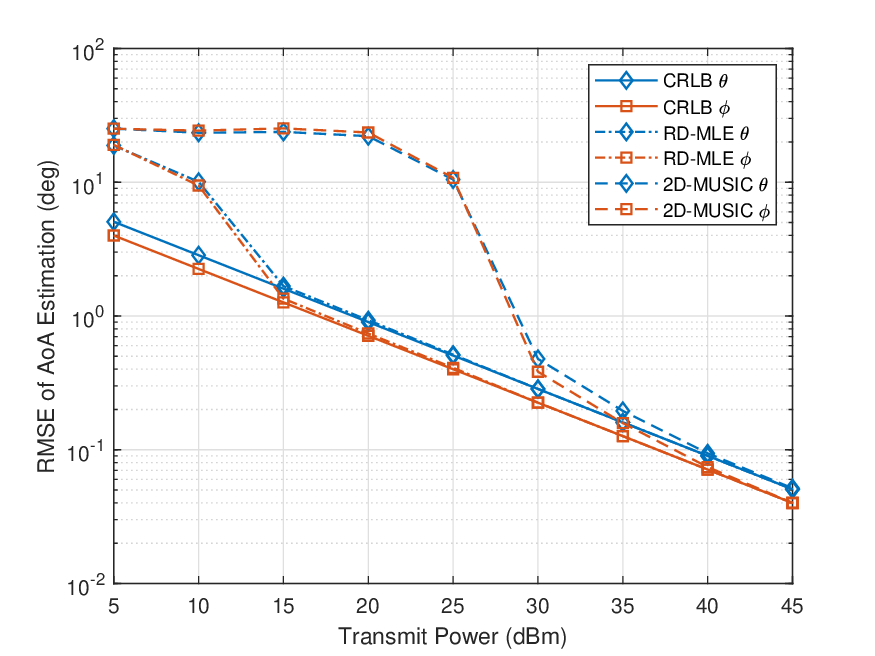}
	\caption{Target angle estimation RMSE under different estimator.}
	\label{CRLB2}
\end{figure}

\begin{figure}[t]
	\centering
	\includegraphics[width=0.8\linewidth]{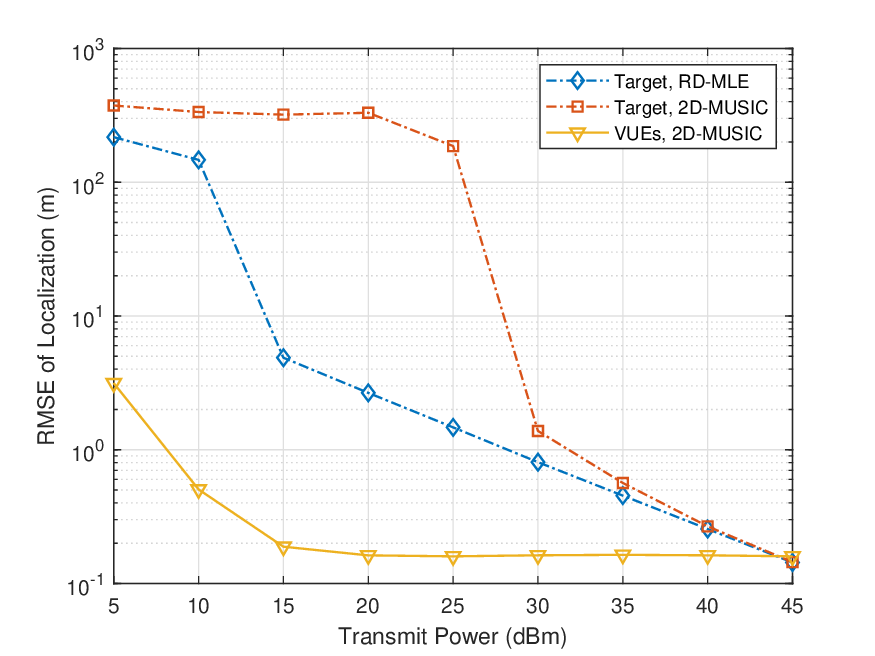}
	\caption{Localization RMSE versus BS's transmit power.}
	\label{rmse_loca}
\end{figure}

In \figurename{ \ref{CRLB2}}, we show the root mean squared error (RMSE) of AoA estimation for the weak target under different estimators. \textcolor{black}{It can be observed that the CRLB serves as a lower bound for the RMSE in angle estimation, and can be approached by both the 2D-MUSIC method and the proposed RD-MLE method. This proves that adopting the CRLB as the performance metric can indeed improve the target estimation performance.} It is worth noting that the proposed RD-MLE method outperforms the 2D-MUSIC method, especially in the low transmit power region. \textcolor{black}{Besides, we present the RMSE of localization based on the 2D AoA in \figurename{ \ref{rmse_loca}}. It is observed that the target can be accurately located in the high transmit power region, where the sensing results are utilized for beamforming design. Moreover, VUEs can be located with lower transmit power due to the larger RCS. Therefore, the LoS component can be easily reconstructed to assist beamforming design. Note that the localization accuracy cannot improve anymore when $p$ exceeds 20 dBm due to the interference between multiple targets.}

\subsection{Communication Performance Evaluation}
\begin{figure}[t]
	\centering
	\includegraphics[width=0.8\linewidth]{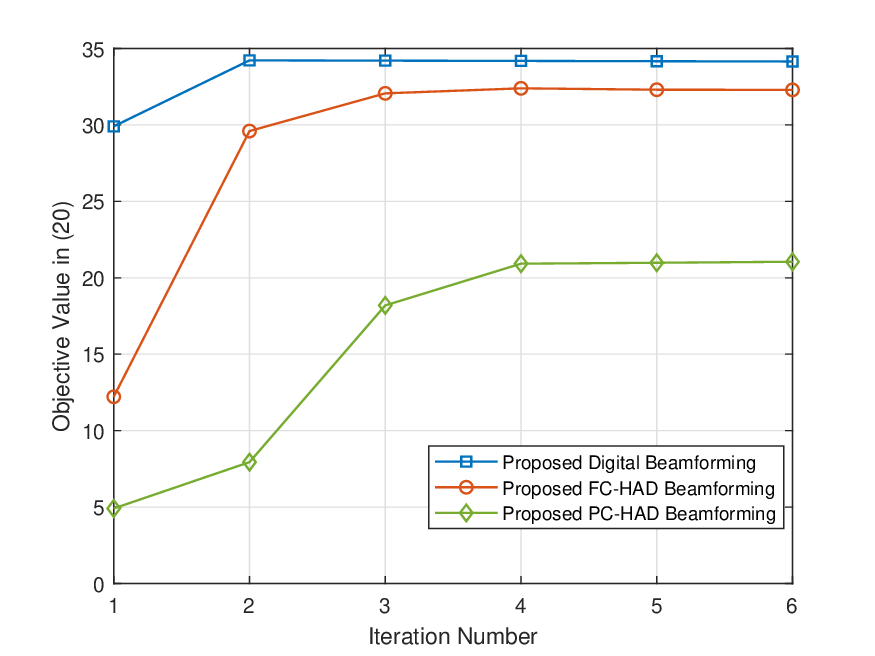}
	\caption{Convergence behaviors of the beamforming design algorithm.}
	\label{converge}
\end{figure}

In this subsection, we evaluate the communication performance of the proposed algorithm for maximizing the sum rate. \textcolor{black}{In \figurename{ \ref{converge}}, we compare the convergence of the proposed PDD-based beamforming design algorithm for fully digital arrays, fully-connected HAD arrays, and partially-connected HAD arrays (labeled with `Proposed Digital Beamforming', `Proposed FC-HAD Beamforming', and  `Proposed PC-HAD Beamforming'), where the outer iterations of the PDD method are presented. As can be seen, the objective value obtained by the proposed algorithm is able to converge in only a few iterations due to both the excellent convergence performance of the PDD method and effective initialization.} Note that the algorithm convergence speed in the fully digital architecture is the fastest, followed by the fully-connected architecture, and the slowest in the partially-connected architecture. Moreover, the fully digital array achieves the highest objective value and the partially-connected array achieves the lowest. The reason is that the optimal solution of \eqref{eq4.9} in each PDD iteration can be obtained. For the hybrid architecture, the digital beamformer and analog beamformer are optimized in an alternative manner, potentially leading to local optima. Moreover, the performance deterioration is exacerbated by limited DoFs inherent with a restricted number of RF chains, especially for the partially-connected architecture. 

\begin{figure}[t]
	\centering
	\includegraphics[width=0.8\linewidth]{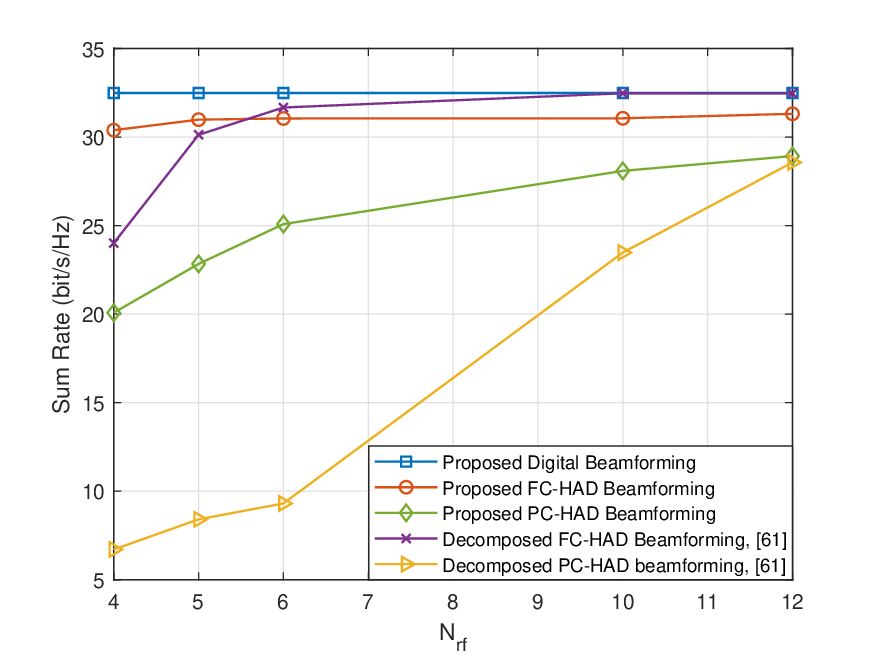}
	\caption{Sum-rate versus number of RF chains, $M=60$.}
	\label{main_Nrf}
\end{figure}

In \figurename{ \ref{main_Nrf}}, we investigate the sum rate of VUEs under different numbers of RF chains. The designed beamforming matrix is applied in Rician channels where the LoS component is reconstructed via sensing results based on \eqref{eq2.2}. For comparison, the performance of the conventional hybrid beamforming algorithm in \cite{7397861} is also provided (for fully-connected HAD array and partially-connected HAD array, the curves are labeled with `Decomposed FC-HAD beamforming' and `Decomposed PC-HAD beamforming', respectively.), where the hybrid beamforming matrices are obtained by decomposing the fully digital beamformer, \textcolor{black}{whose computational complexity $\mathcal{O}\left(n_{iter} M N_{rf} K \right)$, where $n_{iter}$ denotes the iteration number. Despite lower complexity, this method relies on the pre-optimized fully digital beamformer. } It is shown that the fully digital beamformer exhibits the best performance as expected. The Decomposed FC-HAD beamforming scheme achieves the same sum rate as the fully digital beamformer when deploying plenty of RF chains, while the performance severely degrades with fewer RF chains. This is because the optimal solution of a fully-connected HAD beamformer can be obtained when $N_{rf} \ge 2 N_s$ \cite{6966076}. Otherwise, the approximation error is inevitable. In comparison, the fully-connected HAD beamformer obtained by the proposed PDD-based algorithm performs excellently even with few RF chains. It is a pity that the sum rate does not remarkably increase with more RF chains, where the local optima of the alternating algorithm becomes the major performance bottleneck. As for the partially-connected HAD array, the proposed PDD-based algorithm significantly outperforms the decomposition-based algorithm. 

\begin{figure}[t]
	\centering
	\includegraphics[width=0.8\linewidth]{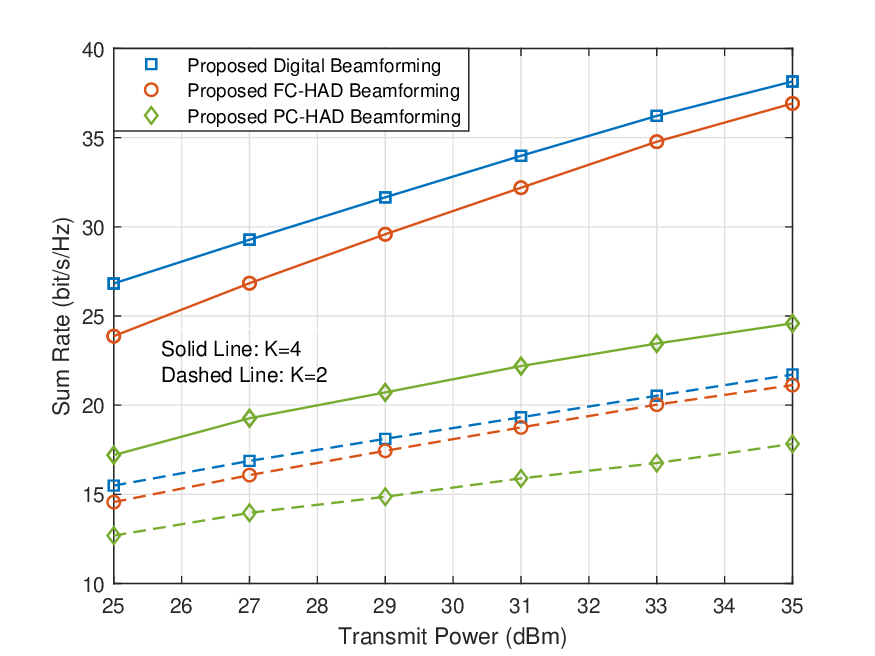}
	\caption{Sum-rate versus BS's transmit power.}
	\label{main_pk}
\end{figure}

\begin{figure}[t]
	\centering
	\includegraphics[width=0.8\linewidth]{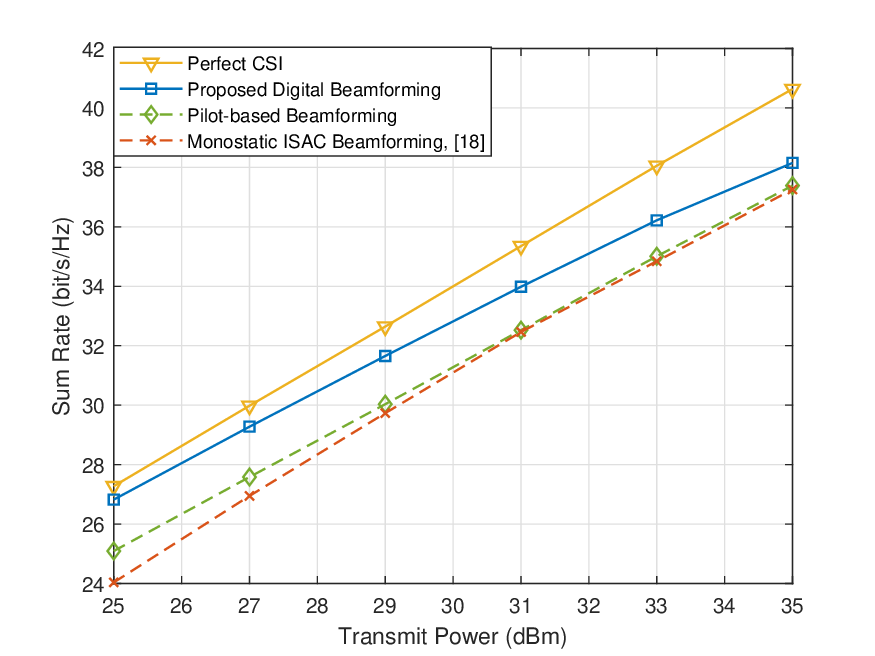}
	\caption{Sum-rate achieved by different beamforming designs.}
	\label{main_pk_compare}
\end{figure}

In \figurename{ \ref{main_pk}}, we present the sum rate versus the BS's transmit power. As can be seen, under both cases of $K=2$ and $K=4$, the sum rate increases with the power budget due to the following two reasons. First, the SINR at each VUE is improved with higher transmit power. Second, the CRLB constraint is easier to satisfy with higher transmit power, thus providing greater DoFs for beamforming optimization.

\textcolor{black}{In \figurename{ \ref{main_pk_compare}}, we compare the performance of the proposed algorithm with different schemes, i.e., the ``Perfect CSI'' scheme, the ``Monostaic ISAC Beamforming" scheme, and the ``Pilot-based Beamforming" scheme. In the ``Perfect CSI'' schemes, the perfect CSI, including both LoS and NLoS components, is always considered available, then we design the corresponding beamforming matrices based on the proposed algorithm, which serves as a performance upper bound. In the ``Pilot-based Beamforming" schemes, the perfect CSI is assumed to be obtained via pilot symbols at the cost of $8\%$ temporal resources \cite{66}, and the beamforming is designed by the proposed algorithm. In the ``Monostaic ISAC Beamforming'' schemes, the algorithm in \cite{10364735} is applied to the V2X scenario, where the perfect CSI is obtained by pilot symbols. First, it can be observed that the proposed algorithm outperforms the ``Pilot-based Beamforming" scheme and the ``Monostaic ISAC Beamforming'' scheme, since the pilot overhead is reduced by the sensing-assisted beamforming design mechanism. Moreover, the performance of ``Pilot-based Beamforming" scheme achieves a higher data rate than the ``Monostaic ISAC Beamforming'' scheme,  especially for the low transmit power region. The reason is that with the aid of sensing node near the target, the CRLB constraint is easier to satisfy, thus reserving more DoFs for beamforming optimization. Also note that, in this paper, only the LoS channel component has been reconstructed, while the NLoS ones are omitted; hence, the increase of transmit power also results in uncontrollable multi-user interference, thereby degrading the increasing speed of sum rate for the ``Proposed Digital Beamforming" scheme.}

\begin{figure}[t]
	\centering
	\includegraphics[width=0.8\linewidth]{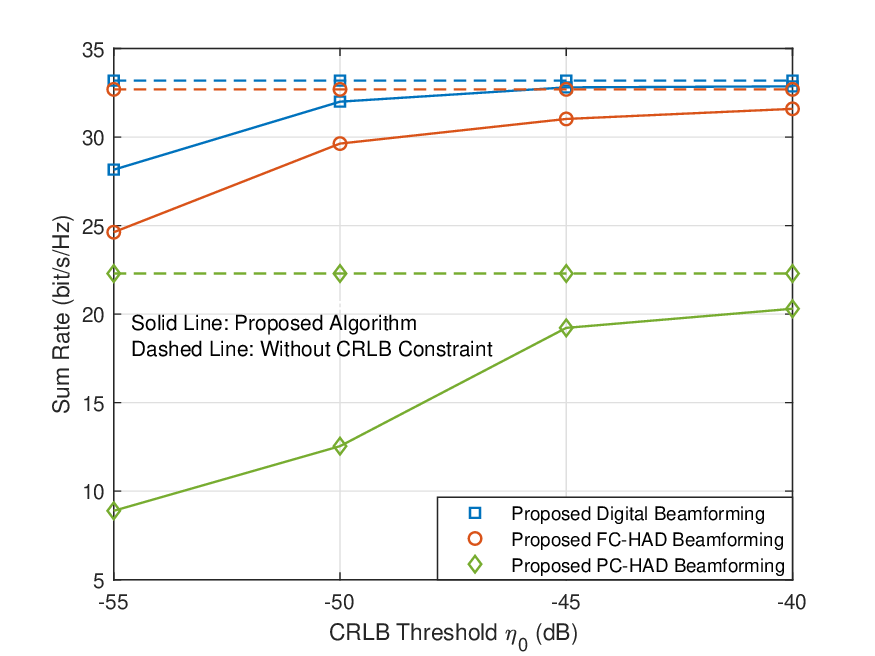}
	\caption{Trade-off between sum rate and CRLB threshold $\eta_0$.}
	\label{main_eta}
\end{figure}

In \figurename{ \ref{main_eta}}, we show the variation of the sum rate with different CRLB constraint thresholds for the weak target to depict the trade-off between the communication performance and sensing performance. \textcolor{black}{With the increase of CRLB threshold, the constraint on the sensing functionality becomes looser, and more power can be radiated towards the directions of VUEs. Therefore, the communication performance is improved and approaches that without CRLB constraint.} An interesting phenomenon is that the identical CRLB constraint is stricter for partially-connected HAD arrays than for fully digital arrays and fully-connected HAD arrays. Specifically, the sum rate of fully digital arrays and fully-connected HAD arrays significantly increases when the CRLB threshold is relaxed to $-50$ dB from $-55$ dB. In comparison, a higher sum rate of partially-connected HAD arrays can only be achieved with a looser constraint than $-45$ dB. This is still due to the limited DoFs of the partially-connected architecture makes it difficult to achieve both high-accuracy sensing and high-speed communication simultaneously.

\section{Conclusion}
In this paper, we investigated beamforming design for a novel SSAC system, where a specialized sensing node was deployed to sense the target and the VUEs based on the echoes of the communication signals transmitted by BS. In addition, the sensing result was utilized to assist in beamforming design, thus improving communication performance. We formulated a sum rate maximization problem under the CRLB constraint for target 2D AoA estimation. To solve the non-convex problem, we proposed an alternating algorithm based on the FP technique and PDD framework for the fully digital array. Then, the proposed PDD-based beamforming design algorithm was extended to the HAD array, where the manifold optimization was employed to tackle the unit-modulus constraint. The effectiveness of the proposed design was verified in terms of convergence, sensing accuracy, and achievable rate performance. Simulation results showed that the proposed SSAC framework can achieve satisfying sensing and communication performance while avoiding the self-interference issue in conventional ISAC systems. \textcolor{black}{In future, we will further investigate the robust beamforming algorithm to address the multi-device interference, which is caused by imperfect CSI reconstructed based on the sensing results. Moreover, we will also investigate the performance gain of incorporating dedicated sensing signals in improving sensing accuracy.}

\section*{Appendix A: Reduced-Dimension Maximum Likelihood Estimate}
In this paper, we assume the sensing node cooperates with the BS with an optical cable connection. Therefore, the transmit communication signal $\bf X$ is deterministic for the sensing node. According to the likelihood function of ${\bf y}_r$ in \eqref{eq2.9add}, the MLE of $\boldsymbol{\xi}$ is given by
\begin{align}
	\label{eqapp1}
	\boldsymbol{ \xi}_{ml}  = \arg \mathop {\max }\limits_{\boldsymbol{\xi}}  p\left( {{{\bf{y}}_r}| {\boldsymbol{\xi}} } \right) = \arg \mathop {\min }\limits_{\boldsymbol{\hat \xi}} {\left\| {{{\bf{y}}_r} - \hat \alpha {\bf \hat g} } \right\|_F^2},
\end{align} 
where ${\bf \hat g} = {\bf g}(\hat \theta_t,\hat \phi_t, \hat \psi_t)$, ${\boldsymbol{\hat \xi}} = \left[\hat \theta_t, \hat \phi_t, \hat \psi_t, \Re(\hat \alpha), \Im(\hat \alpha) \right]^T$ denote the unknown parameters to be estimated. Note that, for any given $\hat \theta_t$, $\hat \phi_t$ and $\hat \psi_t$, the least-squares solution for $\hat \alpha$ is
\begin{align}
	\label{eqapp2}
	\alpha_{ml}  = \arg \mathop {\min }\limits_{\hat \alpha}  {\left\| {{{\bf{y}}_r} - {\hat \alpha} {\bf \hat g} } \right\|_F^2} = \frac{{{{\bf \hat g}^H}}}{{\left\| {\bf \hat g} \right\|_F}}{{\bf{y}}_r}.
\end{align} 
By substituting the $\alpha_{ml}$ into \eqref{eqapp1}, the MLE of $\theta_t$, $\phi_t$ and $\psi_t$ is given as 
\begin{equation}	
	\label{eqapp3}	
	\begin{aligned}	
		\left[ {{\theta_t^{ml}},{\phi_t^{ml}},{\psi_t^{ml}}} \right] &= \arg \mathop {\min }\limits_{{\hat \theta _t},{\hat  \phi _t},{\hat \psi _t}} {\left\| {{{\bf{y}}_r} - {\hat \alpha} {\bf \hat{g}}} \right\|_F^2}\\
		& = \arg \mathop {\min }\limits_{{\hat \theta _t},{\hat \phi _t},{\hat \psi _t}} \left\| {{\bf{y}}_r} \right\|_F^2- \frac{{\left| { {\bf{\hat g}}^H {\bf{y}}_r} \right|}}{{{{\left\| {{\bf{\hat g}}} \right\|_F}^2}}},\\
		& = \arg \mathop {\max }\limits_{{\hat \theta _t},{\hat \phi _t},{\hat \psi _t}} {{\left| { {\bf{\hat g}}^H {\bf{y}}_r} \right|}}.
	\end{aligned} 
\end{equation}
Note that the ${\left\| {{\bf{ \hat g}}} \right\|_F^2} = {\rm tr}\left( {{{\bf{X}}^H}{\bf{X}}} \right) = P$. The MLE of the parameters can be obtained by exhaustively searching. However, the three-dimensional exhaustively searching is computational complexity. Moreover, the interested parameters are the 2D AoAs ($\theta_t$, $\phi_t$), and the estimation for $\psi_t$ is redundant. By substituting \eqref{eq2.9} into \eqref{eqapp2}, the expectation can be expressed as
\begin{equation}	
	\label{eqapp4}	
	\begin{aligned}	
		&\arg \mathop {\max }\limits_{{{\hat \theta }_t},{{\hat \phi }_t},{{\hat \psi }_t}} \mathbb{E} \left( {\left| {{{{\bf{\hat g}}}^H}{{\bf{y}}_r}} \right|} \right)\\
		= &\arg \mathop {\max }\limits_{{{\hat \theta }_t},{{\hat \phi }_t},{{\hat \psi }_t}} \mathbb{E} \left( {\left| {{\rm vec}{{\left( {{{{\bf{\hat a}}}_r}{{{\bf{\hat b}}}^H}{\bf{X}}} \right)}^H} {\rm vec}\left( {\alpha {{\bf{a}}_r}{{\bf{b}}^H}{\bf{X}} + {{\bf{N}}_s}} \right)} \right|} \right)\\
		= &\arg \mathop {\max }\limits_{{{\hat \theta }_t},{{\hat \phi }_t},{{\hat \psi }_t}}  \mathbb{E} \left( {\left| {\alpha {\bf{\hat a}}_r^H{{\bf{a}}_r} {\rm Tr}\left( {{{\bf{X}}^H}{\bf{\hat b}}{{\bf{b}}^H}{\bf{X}}} \right) + {\rm Tr}\left( {{{\bf{X}}^H}{\bf{\hat b\hat a}}_r^H{{\bf{N}}_s}} \right)} \right|} \right),
	\end{aligned} 
\end{equation}
where ${{{\bf{\hat a}}}_r} = {{\bf{a}}_r}\left( {{{\hat \theta }_t},{{\hat \phi }_t}} \right)$, ${\bf{\hat b}} = {\bf{b}}\left( {\hat \psi } \right)$. For any given $\psi_t$, the objective value is maximized when $\hat \theta_t = \theta_t$, $\hat \phi_t = \phi_t$, and the \eqref{eqapp4} can be rewritten as 
\begin{equation}	
	\label{eqapp5}	
	\begin{aligned}	
	\left[ {\theta _t,\phi _t} \right] = \arg \mathop {\max }\limits_{{{\hat \theta }_t},{{\hat \phi }_t}} {\bf{\hat a}}_r^H{{\bf{a}}_r} {\left| {\alpha {\rm tr}\left( {{{\bf{X}}^H}{\bf{\hat b}}{{\bf{b}}^H}{\bf{X}}} \right)} \right|} .
	\end{aligned} 
\end{equation}
Therefore, the MLE can be recast as the reduced dimension form,
\begin{equation}	
	\label{eqapp6}	
	\begin{aligned}	
		\left[ {{\theta_t^{rd}},{\phi_t^{rd}}} \right] = \arg \mathop {\max }\limits_{{\hat \theta _t},{\hat \phi _t}} {{\left| { {\bf{\hat g}}^H(\hat \theta_t,\hat \phi_t, \hat \psi_t) {\bf{y}}_r} \right|}},
	\end{aligned} 
\end{equation}
where the $\hat \psi_t$ denotes any given estimated $\psi_t$, and the sensing performance for $\theta _t$ and $\phi _t$ is not significantly affected by the set of $\hat \psi_t$. In practice, the $\hat \psi_t$ can be set as the value in the previous frame, which can be calculated based on the estimated $\hat \theta_t$ and $\hat \psi_t$ in the previous frame, or just the interested direction for simplicity.

\bibliographystyle{IEEEtran}%
\bibliography{SSAC}

\end{document}